\documentclass[iop,numberedappendix]{emulateapj}

\usepackage{graphicx,amsmath,natbib}
\usepackage{here}
\usepackage{epsf}
\usepackage{epstopdf}
\usepackage{xspace}

\usepackage[normalem]{ulem} 
\usepackage{cancel} 

\usepackage{color}

\newcommand{\um}{\ensuremath{\mu\rm{m}}\xspace}

\newcommand{\lir}{\ensuremath{L_{\rm{IR}}}\xspace}
\newcommand{\msol}{\ensuremath{\rm{M}_\odot}\xspace}
\newcommand{\lsol}{\ensuremath{\rm{L}_\odot}\xspace}
\newcommand{\twco}{\ensuremath{^{12}\rm{CO}}\xspace}
\newcommand{\twc}{\ensuremath{^{12}\rm{C}}\xspace}
\newcommand{\thc}{\ensuremath{^{13}\rm{C}}\xspace}
\newcommand{\eo}{\ensuremath{^{18}\rm{O}}\xspace}
\newcommand{\thco}{\ensuremath{^{13}\rm{CO}}\xspace}
\newcommand{\ceto}{\ensuremath{\rm{C}^{18}\rm{O}}\xspace}
\newcommand{\hcop}{\ensuremath{\rm{HCO}^+}\xspace}
\newcommand{\hto}{\ensuremath{\rm{H}_2\rm{O}}\xspace}

\newcommand{\hctn}{\ensuremath{\rm{HC}_3\rm{N}}\xspace}
\newcommand{\percc}{\ensuremath{\rm{cm}^{-3}}\xspace}
\newcommand{\apm}{APM~0827+5255\xspace}
\newcommand{\sopf}{\ensuremath{S_\mathrm{1.4\,mm}}\xspace}
\newcommand{\sef}{\ensuremath{S_\mathrm{850\,\um}}\xspace}
\newcommand{\smm}{SMM~J2135-0102\xspace}
\newcommand{\tkin}{\ensuremath{T_\mathrm{kin}}\xspace}
\newcommand{\nHt}{\ensuremath{n_\mathrm{H_2}}\xspace}
\newcommand{\ncrit}{\ensuremath{n_\mathrm{crit}\xspace}}
\newcommand{\Ndv}{\ensuremath{\rm{cm}^{-2}(\rm{km}/\rm{s})^{-1}}\xspace}

\def\Arizona{1}
\def\UPenn{2}
\def\ESO{3}
\def\Diego{4}
\def\CfA{5}
\def\Saclay{6}
\def\JPL{7}
\def\Cambridge{8}
\def\Miss{9}
\def\KICPChicago{10}
\def\EFIChicago{11}
\def\PhysicsUChicago{12}
\def\AAUChicago{13}
\def\Dal{14}
\def\ESOGarching{15}
\def\Davis{16}
\def\UFlorida{17}
\def\UCL{18}
\def\McGill{19}
\def\Berkeley{20}
\def\Bristol{21}
\def\UCLA{22}
\def\Carnegie{23}
\def\MPIfR{24}
\def\Caltech{25}
\def\Illinois{26}
\def\APC{27}

\begin{document}

\title{The Rest-Frame Submillimeter Spectrum\\of High-Redshift, Dusty, Star-Forming Galaxies}

\shortauthors{J. S. Spilker, et al.}

\author{
J.~S.~Spilker$^{\Arizona,*}$,
D.~P.~Marrone$^{\Arizona}$,   
J.~E.~Aguirre$^{\UPenn}$,
M.~Aravena$^{\ESO,\Diego}$,
M.~L.~N.~Ashby$^{\CfA}$,
M.~B\'ethermin$^{\Saclay}$,
C.~M.~Bradford\altaffilmark{\JPL},
M.~S.~Bothwell$^{\Cambridge}$,
M.~Brodwin$^{\Miss}$,
J.~E.~Carlstrom$^{\KICPChicago,\EFIChicago,\PhysicsUChicago,\AAUChicago}$, 
S.~C.~Chapman$^{\Dal}$,
T.~M.~Crawford$^{\KICPChicago,\AAUChicago}$, 
C.~de~Breuck$^{\ESOGarching}$,
C.~D.~Fassnacht$^{\Davis}$,
A.~H.~Gonzalez$^{\UFlorida}$, 
T.~R.~Greve$^{\UCL}$,	
B.~Gullberg$^{\ESOGarching}$, 
Y.~Hezaveh$^{\McGill,\dagger}$,
W.~L.~Holzapfel$^{\Berkeley}$, 
K.~Husband$^{\Bristol}$,
J.~Ma$^{\UFlorida}$, 
M.~Malkan$^{\UCLA}$,
E.~J.~Murphy$^{\Carnegie}$,
C.~L.~Reichardt$^{\Berkeley}$, 
K.~M.~Rotermund$^{\Dal}$,
B.~Stalder$^{\CfA}$, 
A.~A.~Stark$^{\CfA}$, 
M.~Strandet$^{\MPIfR}$, 
J.~D.~Vieira$^{\Caltech,\Illinois}$,
A.~Wei\ss$^{\MPIfR}$,
N.~Welikala$^{\APC}$
}

\altaffiltext{\Arizona}{Steward Observatory, University of Arizona, 933 North Cherry Avenue, Tucson, AZ 85721, USA}
\altaffiltext{$*$}{Email address: \href{mailto:jspilker@as.arizona.edu}{jspilker@as.arizona.edu}}
\altaffiltext{\UPenn}{University of Pennsylvania, 209 South 33rd Street, Philadelphia, PA 19104, USA}
\altaffiltext{\ESO}{European Southern Observatory, , Alonso de Cordova 3107, Casilla 19001 Vitacura Santiago, Chile.}
\altaffiltext{\Diego}{N\'ucleo de Astronom\'{\i}a, Facultad de Ingenier\'{\i}a, Universidad Diego Portales, Av. Ej\'ercito 441, Santiago, Chile}
\altaffiltext{\CfA}{Harvard-Smithsonian Center for Astrophysics, 60 Garden Street, Cambridge, MA 02138, USA}
\altaffiltext{\Saclay}{Laboratoire AIM-Paris-Saclay, CEA/DSM/Irfu - CNRS - Universit\'e Paris Diderot, CEA-Saclay, Orme des Merisiers, F-91191 Gif-sur-Yvette, France}
\altaffiltext{\JPL}{Jet Propulsion Laboratory, 4800 Oak Grove Drive, Pasadena, CA 91109, USA}
\altaffiltext{\Cambridge}{Cavendish Laboratory, University of Cambridge, JJ Thompson Ave, Cambridge CB3 0HA, UK}
\altaffiltext{\Miss}{Department of Physics and Astronomy, University of Missouri, 5110 Rockhill Road, Kansas City, MO 64110, USA}
\altaffiltext{\KICPChicago}{Kavli Institute for Cosmological Physics, University of Chicago, 5640 South Ellis Avenue, Chicago, IL 60637, USA}
\altaffiltext{\EFIChicago}{Enrico Fermi Institute, University of Chicago, 5640 South Ellis Avenue, Chicago, IL 60637, USA}
\altaffiltext{\PhysicsUChicago}{Department of Physics, University of Chicago, 5640 South Ellis Avenue, Chicago, IL 60637, USA}
\altaffiltext{\AAUChicago}{Department of Astronomy and Astrophysics, University of Chicago, 5640 South Ellis Avenue, Chicago, IL 60637, USA}
\altaffiltext{\Dal}{Dalhousie University, Halifax, Nova Scotia, Canada}
\altaffiltext{\ESOGarching}{European Southern Observatory, Karl Schwarzschild Stra\ss e 2, 85748 Garching, Germany}
\altaffiltext{\Davis}{Department of Physics,  University of California, One Shields Avenue, Davis, CA 95616, USA}
\altaffiltext{\UFlorida}{Department of Astronomy, University of Florida, Gainesville, FL 32611, USA}
\altaffiltext{\UCL}{Department of Physics and Astronomy, University College London, Gower Street, London WC1E 6BT, UK}
\altaffiltext{\McGill}{Department of Physics, McGill University, 3600 Rue University, Montreal, Quebec H3A 2T8, Canada}
\altaffiltext{\Berkeley}{Department of Physics, University of California, Berkeley, CA 94720, USA}
\altaffiltext{\Bristol}{H.H. Wills Physics Laboratory, University of Bristol, Tyndall Avenue, Bristol BS8 1TL, UK}
\altaffiltext{\UCLA}{Department of Physics and Astronomy, University of California, Los Angeles, CA 90095-1547, USA}
\altaffiltext{\Carnegie}{Observatories of the Carnegie Institution for Science, 813 Santa Barbara Street, Pasadena, CA 91101, USA}
\altaffiltext{\MPIfR}{Max-Planck-Institut f\"{u}r Radioastronomie, Auf dem H\"{u}gel 69 D-53121 Bonn, Germany}
\altaffiltext{\Caltech}{California Institute of Technology, 1200 E. California Blvd., Pasadena, CA 91125, USA}
\altaffiltext{\Illinois}{Department of Astronomy and Department of Physics, University of Illinois, 1002 West Green Street, Urbana, IL 61801, USA}
\altaffiltext{\APC}{AstroParticule et Cosmologie, Universit\'{e} Paris Diderot, CNRS/IN2P3, CEA/lrfu, Observatoire de Paris, Sorbonne Paris Cit\'{e}, 10, rue Alice Domon et L\'{e}onie Duquet, 75205 Paris Cedex 13, France}
\altaffiltext{\textdagger}{Current Address: Kavli Institute for Particle Astrophysics and Cosmology, Stanford University, Stanford, CA 94305, USA}

\begin{abstract}

We present the average rest-frame spectrum of high-redshift dusty, star-forming galaxies from ${250-770}$\,GHz.  This spectrum was constructed by stacking ALMA 3\,mm spectra of 22 such sources discovered by the South Pole Telescope and spanning $z=2.0-5.7$.  In addition to multiple bright spectral features of \twco, [CI], and \hto, we also detect several faint transitions of \thco, HCN, HNC, \hcop, and CN, and use the observed line strengths to characterize the typical properties of the interstellar medium of these high-redshift starburst galaxies.  We find that the \thco brightness in these objects is comparable to that of the only other $z > 2$ star-forming galaxy in which \thco has been observed.  We  show that the emission from the high-critical density molecules HCN, HNC, \hcop, and CN is consistent with a warm, dense medium with $\tkin \sim 55$\,K and $\nHt \gtrsim 10^{5.5}$\,\percc.  High molecular hydrogen densities are required to reproduce the observed line ratios, and we demonstrate that alternatives to purely collisional excitation are unlikely to be significant for the bulk of these systems.  We quantify the average emission from several species with no individually detected transitions, and find emission from the hydride CH and the linear molecule CCH for the first time at high redshift, indicating that these molecules may be powerful probes of interstellar chemistry in high-redshift systems.  These observations represent the first constraints on many molecular species with rest-frame transitions from $0.4-1.2$\,mm in star-forming systems at high redshift, and will be invaluable in making effective use of ALMA in full science operations.

\end{abstract}


\section{Introduction} \label{intro}
High redshift, dusty, star-forming galaxies (DSFGs) are a population of luminous ($\lir > 10^{12}\lsol$), dust-obscured objects undergoing short-lived intense starburst events \citep[e.g.,][]{blain02,lagache05}.  First discovered by the SCUBA instrument on the James Clerk Maxwell Telescope at 850\,\um in the late 1990s \citep{smail97,barger98,hughes98}, these distant sources are sufficiently faint to make follow-up study at all wavelengths difficult. Additionally, the large beam sizes of single-dish submillimeter facilites has made the identification of optical or infrared counterparts to the submillimeter sources challenging.  Their infrared luminosities imply star formation rates of hundreds to thousands of solar masses per year, making them capable of becoming massive, quiescent galaxies ($M_{*} \sim 10^{11}\msol$) in only 100\,Myr \citep{hainline11,michalowski12,fu13}.  The space and redshift distributions of these extreme starbursts are clearly important diagnostics of the buildup of structure in the universe, but remain a challenge for current galaxy evolution models \citep[e.g.,][]{baugh05,swinbank08,dave10,hayward13}.  In recent years, a picture has emerged in which the majority of gas-rich galaxies lie along a so-called `main sequence' in stellar mass vs. star formation rate, characterized by star formation in massive, secular disks \citep[e.g.,][]{noeske07b,daddi10,tacconi10,elbaz11,hodge12}.  A minority of objects exhibit significantly enhanced star formation rates, and are characterized by star formation triggered by major mergers \citep[e.g.,][]{narayanan09,engel10}.

Given the challenging nature of follow-up observations, the study of gravitationally lensed starburst systems continues to generate valuable insight into the properties and physics of high-redshift DSFGs.  Strong gravitational lensing creates gains in sensitivity or angular resolution which allow much more detailed studies than are possible for otherwise equivalent unlensed systems.  Unfortunately, the brightest sub-mm sources have such low number density ($N < 1$\,$\mathrm{deg}^{-2}$ for $\sef > 100$\,mJy; \citealt{negrello07}) that large area surveys are the only way to build up a statistically significant sample.  Large numbers of such objects have recently been uncovered by wide-field sub/millimeter surveys, including those conducted by the South Pole Telescope (SPT; \citealt{carlstrom11,vieira10,mocanu13}) and \textit{Herschel}/SPIRE \citep{negrello10,wardlow13}.  High-resolution follow-up imaging at 870\,\um has confirmed that these objects are nearly all lensed \citep{hezaveh13,vieira13,bussmann13}.  Lensed DSFGs offer the best chance to observe these systems for spectral lines which would otherwise be too faint to detect at such great distances, allowing a more detailed characterization of the interstellar medium in these objects.

The prodigious star formation rates of DSFGs require that they contain vast reservoirs of molecular gas ($M_{H_2} \sim 10^{10}\msol$; e.g., \citealt{greve05,bothwell13}) from which those stars form.  Probing the density, thermodynamic state, and balance of heating and cooling of the interstellar gas then reveals the star-forming conditions in these extreme starbursts.  Unfortunately, due to its low mass and lack of a permanent electric dipole moment, direct observations of cold H$_2$ are difficult.  Instead, a suite of molecular and atomic fine structure lines are typically used to diagnose the interstellar medium of galaxies both locally and at high redshift.  Carbon monoxide (\twco) is by far the most common molecule observed at millimeter wavelengths in extragalactic objects, due to its high abundance relative to H$_2$, ease of excitation, and rotational lines at frequencies of high atmospheric transmission.  The ground state rotational line of \twco(1-0) ($\nu_\mathrm{rest} = $115\,GHz) has been used for decades \citep[e.g.,][]{wilson70} as a tracer of the bulk of the molecular gas in the interstellar medium.  However, the numerical conversion between gas mass and \twco luminosity can vary by more than an order of magnitude depending on the metallicity and gas conditions of the galaxy, and the appropriate value for most high-redshift systems is uncertain \citep[e.g.,][]{downes98,tacconi08,ivison11,narayanan12,bolatto13}.  Additional consideration of optically thin species, such as \thco and \ceto, may allow for accurate gas mass estimates, if the relative abundances of those species can be estimated.

Due to its low dipole moment ($\sim$0.15\,D), \twco rapidly becomes collisionally thermalized at densities of just $\nHt \sim \rm{few} \times 10^{2}$\,\percc.  Spectral features of other molecules with higher dipole moments, such as HCN, HNC, and \hcop, are thought to arise from regions with higher densities ($\nHt \gtrsim 10^{4}$\,\percc) where stars are actively forming \citep{gao04b}.  The extreme conditions required for these molecules to be collisionally excited, combined with abundances lower than that of \twco by multiple orders of magnitude \cite[e.g.,][]{wang04,martin06}, make their lines faint and observation difficult.  

The extremely wide spectral range and high sensitivity of the \textit{Herschel}/SPIRE-FTS instrument \citep{griffin10} have allowed for spectral observations of nearby Ultra-Luminous Infrared Galaxies (ULIRGs) over the entire far-IR wavelength range.  In the prototypical ULIRG Arp~220, for example, the \twco spectral line energy distribution (SLED) is now complete up to $J=13-12$, and dozens of lines of species including [CI], \hto, HCN, and OH and their ions and/or isotopologues have been seen in emission and absorption \citep{rangwala11,gonzalezalfonso12}.  With a sufficiently wide range of transitions observed, some degeneracies inherent in excitation modeling can be eliminated, and simple geometric models can be constructed to reproduce all the observed spectral features.  Many of these lines can only be observed in local sources from space, making direct comparison between local starbursts and their high-$z$ counterparts challenging.

At high redshift ($z \gtrsim 1$), observations of \twco and various far-IR fine structure lines have become increasingly common (for a recent review, see \citealt{carilli13}), with well-sampled CO SLEDs available for an increasing number of objects \citep[e.g.,][]{weiss07,bradford09,riechers13}.  Observations of other molecular species, on the other hand, remain rare due to the faintness of their lines.  Thus far, detections of fainter molecular lines have been largely confined to extraordinarily luminous, highly gravitationally magnified quasar host galaxies, and only four objects have been detected in multiple molecules or isotopes besides \twco:  H1413+117 (the ``Cloverleaf'' quasar), \apm, a highly magnified quasar host, \smm (the ``Cosmic Eyelash''), a cluster-lensed ULIRG, and HFLS\,3, a \textit{Herschel}-selected starburst at $z=6.3$.  Specific observations of these objects will be discussed in more detail below.

Observations of the interstellar medium of high-$z$ galaxies are being revolutionized with the beginning of science operations by the Atacama Large Millimeter/submillimeter Array (ALMA).  In particular, ALMA has already been used in Cycle~0 to conduct a blind \twco-based redshift survey of 26 high-$z$ star-forming galaxies \citep{vieira13,weiss13}, with spectral features seen in $\sim90$\% of the sample.  Such redshift searches operate by scanning through large swaths of frequency space looking for bright lines of \twco, [CI], and/or \hto.  As a byproduct, they also offer the opportunity to detect emission from a variety of species whose transitions lie in and amongst the brighter lines.  In contrast to previous, narrow-bandwidth targeted studies of specific transitions, blind redshift searches offer information on \textit{all} transitions which fall within the rest-frame frequency range observed, allowing future follow-up observations to focus on detectable species.

Here, we present the detection and analysis of several lines of \thco, HCN, HNC, \hcop, and the CN radical in a stacked spectrum of 22 gravitationally lensed DSFGs spanning $z = 2 - 5.7$ discovered by the SPT.  The stacked spectrum was created utilizing the ALMA~3\,mm spectra obtained as part of the blind redshift search presented in \citet{weiss13}, and spans 250--770\,GHz (0.39--1.2\,mm) in the rest frame.  This stacked spectrum represents a first attempt at quantifying the relative strengths of a host of faint lines in high-redshift DSFGs and addresses the typical ISM conditions which give rise to such lines.

The paper is organized as follows: in \S\ref{obs}, we briefly describe the sample selection and observations used in the construction of the stacked spectrum.  In \S\ref{stacking}, we describe the method used to scale and stack the spectra of individual objects.  We present the combined spectrum and analyze the average conditions of the ISM in these objects in \S\ref{results}, and conclude by comparing our derived properties to those of other high-redshift systems, constraining the average emission from individually undetected molecules, and discussing alternatives to pure collisional excitation in \S\ref{discussion}.  Throughout this work we adopt a WMAP9 cosmology, with ($\Omega_m, \; \Omega_{\Lambda},\; \mathrm{H}_0) = (0.286,\; 0.713,\; 69.3$ km\,s$^{-1}$\,Mpc$^{-1}$) \citep{hinshaw12}.

\section{Data and Observations} \label{obs}
Extensive details of our target selection and ALMA 3\,mm observations are given in \citet{weiss13}.  Briefly, we selected 26 bright ($S_{1.4\rm{mm}} \gtrsim 20$\,mJy) point sources from the first 1300\,deg$^2$ of the SPT Sunyaev-Zel'dovich effect survey \citep{vieira10,mocanu13}.  These sources showed thermal, dust-like spectral indices between 1.4 and 2\,mm and had no counterparts in existing radio and/or far-IR catalogs, which ruled out synchrotron-dominated sources and low-redshift contaminants.  The sources were required to have been detected with the Large Apex BOlometer CAmera (LABOCA) at 870\,\um or the Submillimeter Array (SMA) at 1.3\,mm to refine their positions.  Due to their extreme brightness, most of the sources were suspected to be gravitationally lensed by intervening massive galaxies, groups, or clusters \citep{negrello07}.  This hypothesis was confirmed using high-resolution ALMA 870\,\um observations \citep{hezaveh13,vieira13}, which show magnification factors $\mu = 5 - 20$.  The sample is not strictly 1.4\,mm flux density limited due to observational constraints, but it does constitute a representative sample of SPT sources which meet the selection criteria.  The sources span redshift $z = 2.01 - 5.70$, apparent infrared luminosity $\lir = 5.5 - 158 \times 10^{12}$\,\lsol (integrated from $8-1000$\,\um), and dust temperature $T_d = 20 - 50$\,K, with medians of $\left\langle z \right\rangle = 3.5$, $\left\langle \lir \right\rangle = 4.2 \times 10^{13}\lsol$, and $\left\langle T_d \right\rangle = 37$\,K.  

The ALMA observations were carried out during Cycle~0 in November 2011 and January 2012 with a compact configuration of the 14--17 antennas available at the time.  The data comprise a spectral sweep of the 3\,mm band (Band 3; \citealt{claude08}), using five tunings of the ALMA receivers to cover 84.2-114.9\,GHz.  Each target was observed for roughly 120\,s per tuning, or a total of $\sim$10\,min per source.  Data were calibrated in the standard way using the Common Astronomy Software Applications package \citep{mcmullin07,petry12}.  Of the 26 targets, redshifts for 18 were unambiguously determined either from the ALMA data alone or in conjunction with additional observations from APEX, ATCA, or VLT; an additional 5 showed a single spectral feature, narrowing the redshift to two or three possible options.  The final three sources showed no lines in their 3\,mm spectra.  

Following publication of \citet{weiss13}, two additional single-line sources have had their redshifts confirmed -- the redshift of SPT\,0125-50 has been confirmed at $z=3.9592$ by the detection of the ground state 1670\,GHz water transition at 336.8\,GHz in high-resolution ALMA 870\,\um data (Appendix \ref{appendix}), and the redshift of SPT\,0512-59 has been confirmed as $z=2.2334$ by the detection of [CII] by \textit{Herschel}/SPIRE (Gullberg et~al., in prep.), leaving just three sources with mulitple redshift options and three sources with no redshift constraint.  For the purposes of this stacking analysis, we include all sources with confirmed redshift, as well as those sources for which a single redshift option is more than 60\% probable (\citealt{weiss13}, their Fig.~4).  The inclusion or exclusion of the sources with ambiguous redshift does not significantly affect our conclusions, and their inclusion is not required for the detection of any faint spectral features.  Conversely, the stacked spectrum and ALMA spectra of the single-line sources are not yet sensitive enough to constrain the redshifts of these sources using the combined constraints of many faint, undetected lines.  The full list of sources included in this study is given in Table~\ref{tab:sources}.

In order to stack and combine the spectra from multiple sources in a statistically robust manner, we must retain an estimate of the flux density uncertainty in each channel of the source spectra.  We fit a point source to the calibrated visibilities of each channel of each source, using the uncertainty from this fitting procedure as an estimate of the noise in each channel.  We see no signs of non-point-like structure in the calibrated continuum images, as expected at the spatial resolution ($>4''$) of the ALMA 3\,mm data.  Since the sources are generally undetected in any given single channel, we fix the source position to that determined from the continuum image of all channels, where every source is robustly detected.  Typical noise levels are $\sim$1.5\,mJy in 45\,MHz channels, increasing to $\sim$2.5\,mJy at the high frequency edge of the ALMA bandpass as the atmospheric transmission declines.   Noise levels are roughly $\sqrt{2}$ lower in the frequency range from 96.2-102.8\,GHz, which was covered twice in our tuning scheme.  The data were acquired with the ALMA correlator configured for 3840 channels per 2\,GHz baseband.  The resulting spectral resolution, $\sim$1.5\,km/s, is much higher than a typical galaxy line width ($\sim$ hundreds of km/s), allowing us to average channels to increase significance.

\begin{deluxetable}{lcccc} 
\tablecaption{Source Properties \label{tab:sources}} 
\startdata 
\tableline 
\\ 
Source Name & z & T$_d$ & Scaled S$_{350\,\mu m}$ & Scaled S$_{21\,\mu m}$ \\ 
 & & (K) & (mJy) & (mJy) \\ 
\tableline 
\\ 
SPT0452-50 & 2.0104 & 21 & 11.7 & 0.0 \\ 
SPT0551-50 & 2.1232 & 27 & 18.2 & 0.1 \\ 
SPT0512-59 & 2.2334$^a$ & 34 & 19.5 & 0.8 \\ 
SPT0125-47 & 2.5148 & 41 & 33.5 & 13.2 \\ 
SPT2134-50 & 2.7799 & 41 & 18.8 & 9.9 \\ 
SPT0103-45 & 3.0917 & 34 & 22.2 & 0.1 \\ 
SPT0550-53 & 3.1286$^b$ & 32 & 9.0 & 1.1 \\ 
SPT0529-54 & 3.3689 & 32 & 20.7 & 0.1 \\ 
SPT0532-50 & 3.3988 & 36 & 34.0 & 2.9 \\ 
SPT0300-46 & 3.5956$^b$ & 39 & 12.3 & 0.4 \\ 
SPT2147-50 & 3.7602 & 42 & 9.4 & 1.4 \\ 
SPT0125-50 & 3.9592$^a$ & 44 & 16.5 & 17.7 \\ 
SPT0418-47 & 4.2248 & 54 & 9.6 & 6.6 \\ 
SPT0113-46 & 4.2328 & 33 & 13.2 & 0.0 \\ 
SPT0345-47 & 4.2958 & 53 & 13.5 & 33.9 \\ 
SPT2103-60 & 4.4357 & 39 & 10.7 & 0.7 \\ 
SPT0441-46 & 4.4771 & 40 & 11.8 & 2.1 \\ 
SPT2146-55 & 4.5672 & 40 & 10.0 & 3.6 \\ 
SPT2132-58 & 4.7677 & 39 & 11.3 & 3.7 \\ 
SPT0459-59 & 4.7993 & 37 & 10.1 & 3.6 \\ 
SPT0346-52 & 5.6559 & 53 & 17.3 & 35.9 \\ 
SPT0243-49 & 5.6990 & 32 & 18.1 & 1.1 \\ 
\enddata 
\tablecomments{Fluxes derived by scaling continuum SEDs for each source to $z=3$ using Equation~\ref{eq:scale} (\S\ref{stacking}) before measuring the flux at rest-frame 350 and 21\,\um (observed-frame 1.4\,mm and 84\,\um).  The observed-frame 1.4\,mm fluxes are used to scale the spectra of individual sources; 21\,\um fluxes will be used in \S\ref{highex} to test the effect of the mid-infrared radiation field on the observed line ratios.}  
\tablenotetext{a}{Redshift confirmed since publication of \citet{weiss13}} 
\tablenotetext{b}{Most likely redshift of two possible options} 
\end{deluxetable}

\section{Stacking Methods} \label{stacking}
The composite spectrum is constructed from individual source spectra that are shifted in redshift and scaled to a common millimeter flux density. This is not the scaling that optimizes the S/N of the stacked spectrum, but it provides a more representative average of the wide distribution of line strengths seen in individual sources at a small cost in the final noise level.  

To create the average DSFG spectrum, we begin by removing a first-order polynomial continuum from the spectrum of each source, excluding channels with significant line emission (defined as channels with signal-to-noise ratios $S/N > 3$).  We then re-scale all spectra to a common redshift, $z_\mathrm{common} = 3.0$.  The multiplicative scaling factor to the channel flux densities and uncertainties of each source which preserves the luminosity per unit bandwidth is 
\begin{equation} \label{eq:scale}
\left(\frac{D_L(z_\mathrm{source})}{D_L(z_\mathrm{common})}\right) ^2 \frac{1 + z_\mathrm{common}}{1+z_\mathrm{source}},
\end{equation}
where $D_L$ is the luminosity distance to the original and common redshifts of each source.  The choice to scale to $z=3$ is intended to be representative of the typical redshift of DSFGs --  \citet{chapman05} find $\left\langle z \right\rangle = 2.3$, while the redshift distribution of the sources presented here implies $\left\langle z \right\rangle = 3.5$.  

We also normalize the flux density and uncertainty of each source by the 1.4\,mm flux density it would have were it located at $z_\mathrm{common}$.  Because the sources have are being shifted to $z=3$, the 1.4~mm flux densities are taken from fits to the continuum SED of each source.  The dust continuum is represented by a single-temperature modified blackbody, with $\beta=2$ and $\lambda_0$, the wavelength at which the dust opacity is unity, fixed to 100\,\um \citep{draine06}.  Additionally, we exclude photometric points at $\lambda_{\mathrm{rest}} < 50$\,\um, as in \citet{greve12}, because the single-temperature SED fits we use below are inappropriate at such short wavelengths, where relatively warm dust contributes significantly.  
The ALMA spectra are then multiplied by the factor that converts the fitted 1.4~mm (rest-frame 350\,\um) flux density to a common $\sopf = 15$\,mJy, 
approximately the average brightness of the SPT sources were they all located at $z=3$.  After scaling each source to $z=3$, DSFGs with scaled $\sopf = 15$\,mJy will have their ALMA flux densities unchanged, while sources fainter (brighter) than this will have their flux densities scaled proportionally higher (lower).  Due to the well-known submillimeter ``negative $k$-correction'' \citep{blain93}, the SPT observed and redshift-scaled 1.4\,mm flux densities are nearly identical for most sources.  The sources with the largest difference between observed and scaled 1.4\,mm flux density are those located at the highest redshifts, where the dust opacity even at observed-frame 1.4\,mm is no longer negligible.  It is these sources which lower the average scaled 1.4\,mm flux density to $\sim$15\,mJy from the observed average of $\sim$20\,mJy.  The choice of ``reference'' wavelength has little effect on the results of our stacking; nearly identical results are obtained referencing to 850\,\um in analogy to samples discovered by SCUBA or LABOCA.

The decision to normalize the ALMA spectra by a scaled version of the sources' milllimeter flux density is by no means unique -- a variety of source properties could conceivably be used for normalization, including apparent \lir, \twco line luminosity, or dust mass.  The line ratios derived from these various weighting schemes are relatively robust to which quantity is used for normalization, with the majority of average line ratios agreeing to within 15\%  regardless of normalization choice.  We choose to normalize by the scaled 1.4\,mm flux density of each source as the SPT sources were selected at this wavelength. 

To create a composite ``template'' spectrum of the SPT DSFGs, we interpolate the spectrum of each source onto a grid spanning 240--780\,GHz, the rest frequencies probed by the ALMA spectra given the redshift distribution of our sources ($z = 2.01 - 5.70$).  The grid is constructed with 500\,km/s spacing, which roughly corresponds to the typical full-width at half-maximum (FWHM) of the observed \twco lines.  We then perform a weighted average of all sources which contribute to a given output channel, with weights determined as the inverse variance in each channel.  Due to the changing number of sources which contribute at each rest frequency (ranging from two to 13), the re-scaling of each spectrum, and the noise properties of the original spectra, the noise level of the composite spectrum varies from approximately 0.11 to 1.5\,mJy in each 500\,km/s output channel, except in the extreme edges of the band where single sources alone remain (see Figure~\ref{fig:composite}).

The template spectrum is an imperfect tool for the detection of faint lines because its regular channel grid necessarily splits  
the flux of most lines between multiple channels, reducing their significance. A cleaner assessment of the presence or absence of spectral features 
can be obtained by constructing spectra centered at the rest frequencies of potentially-detectible ISM lines.
To construct these spectra, we extract the spectrum of every source that samples the line of interest in very wide 600\,km/s channels centered on the target line.
A constant continuum level is estimated and subtracted from each source before stacking, excluding the central 0\,km/s channel and any other channels which also potentially contain other faint lines.  The wide 600\,km/s channel width was chosen to match the typical \twco line widths of our sources, $\left\langle \rm{FWHM} \right\rangle \sim$ 450\,km/s.  For a perfectly Gaussian line of $\mathrm{FWHM} = 450$\,km/s, a channel of 600\,km/s centered at the rest frequency of the line contains $\gtrsim90\%$ of the total line flux.  This choice assumes that the widths of all lines will be roughly equivalent to the widths of \twco, but faint transitions, especially those with high critical densities, frequently show narrower line widths than \twco \citep[e.g.,][]{weiss07}.  With 600\,km/s channelization, therefore, essentially all flux from narrower lines will be contained within a single channel.  The significance of the detection or non-detection of each line can then be interpreted easily based on the flux and noise properties of the central 0\,km/s channel.  When applied to the stacked bright \twco lines themselves, this channelization indeed recovers $\sim90\%$ of the total integrated flux in the channel centered on the rest frequency of each line, suggesting that this chosen width is an acceptable compromise between including as much line flux as possible while excluding excess noise.  We classify a line as detected if the line flux in the 0\,km/s channel has $S/N > 3$.

This procedure is complicated by those species with fine and/or hyperfine structure components which are too close together to be separated, the most relevant of which is the cyanide radical CN.  For these molecules, we choose the main fine structure component as the rest frequency of the line, using spectroscopic data from the JPL \citep{pickett98} and CDMS \citep{muller01,muller05} line catalogs.  CN presents an additional complication in its $N = 4-3$ lines, the strongest components of which are separated by less than a typical FWHM ($\sim$ 90 and 220\,km/s) from the nearby HNC($J=5-4$) line.  These lines are blended in two adjacent 600\,km/s channels.  To attempt to separate them, we use the line ratio derived from the only high-redshift observation of this pair of lines, in the \apm quasar \citep{guelin07}.  We assign line flux in accordance with that ratio (36\% and 64\% for CN(4--3) and HNC(5--4), respectively), and include an additional 33\% uncertainty in the derived line fluxes.

\begin{figure}[htb]%
\includegraphics[width=\columnwidth]{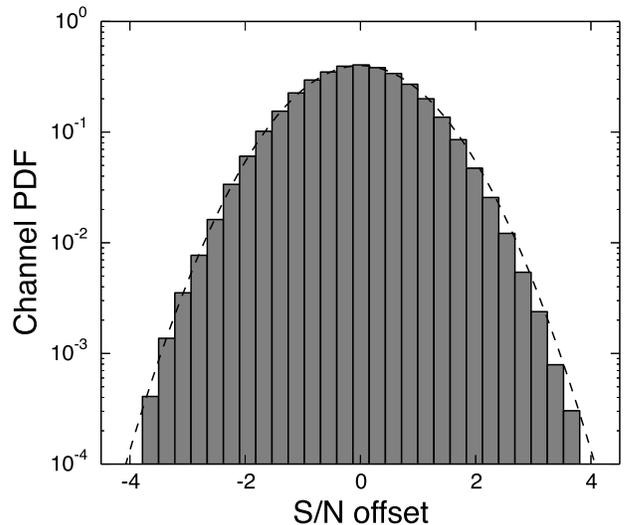}%
\caption{Testing the statistical properties of the stacking procedure.  Shown is the signal-to-noise distribution of channels in the stacked spectrum near rest-frame 513\,GHz, where no line is expected. The channels of the individual source spectra included in the stack were randomized before stacking, 3000 trials were used.  The dashed line indicates a unit-width Gaussian distribution.  The normal distribution of these channels indicates that the statistics of the stack are well-described by a Gaussian distribution with the width specified by the assumed noise level.}%
\label{fig:noisetest}%
\end{figure}

To ensure that the stacking method does not introduce a spurious line where none exists, we use a Monte Carlo procedure to characterize the statistical properties of the stacked spectrum at a line-free frequency, 513\,GHz.  For each source that covers this rest-frame frequency, we extract spectra in 600\,km/s channels within $\pm$4500\,km/s.  In 3000 trials, we randomly shuffle the channels of each input spectrum, then scale, normalize, and average.  The flux in each channel is not altered, but the placement of the channels is randomized such that emission originally located in the 0\,km/s channel may now be located in any other channel.  We then stack the spectra as usual and bin the signal-to-noise (S/N) distribution of the resulting stacked channels from the entire $\pm$4500\,km/s velocity range.  This channel distribution is shown in Fig.~\ref{fig:noisetest}, which indicates that the noise properties of the stacked spectra are well-represented by Gaussian statistics.  The probability of  falsely finding a stacked line with $\left| \mathrm{S/N} \right|$ $>$ 3 is therefore very small; we expect $<0.3$ false detections in the $\sim$100 line frequencies we search.

\section{Results} \label{results}
\begin{figure*}[htb]%
\centering
\includegraphics[width=\textwidth]{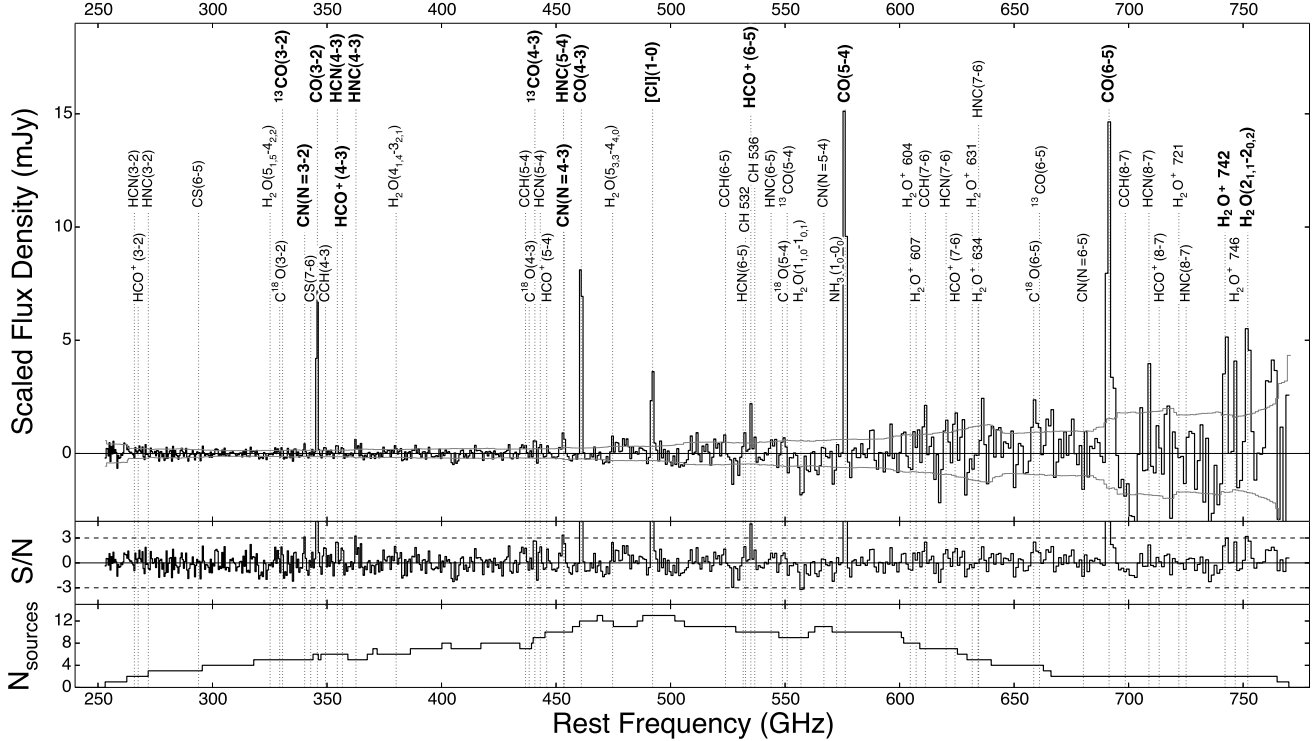}%
\caption{The composite continuum-subtracted rest-frame $0.4-1.2$\,mm spectrum of high-redshift submillimeter galaxies, constructed from 22 SPT DSFGs and shown at 500\,km/s resolution.  A selection of potentially-detectable molecular lines are marked.  Lines we detect using the stacking procedure detailed in \S\ref{stacking} are labeled in large font, and the running $\pm1\sigma$ noise level is shown in grey.  The middle panel shows the running signal-to-noise ratio of the top panel, while the bottom panel shows the number of sources which contribute at each frequency.}%
\label{fig:composite}%
\end{figure*}

We show the average 250--770\,GHz spectrum of the SPT DSFGs in Fig.~\ref{fig:composite}, with a sample of detected and potentially-detectable ISM lines marked.  A summary of the detected lines and an assortment of upper limits is given in Table~\ref{tab:lines}.  This table is not intended to be an exhaustive list of useful ISM diagnostics, but rather gives a host of lines in our bandpass that may be detected in high-redshift starbursts by ALMA in the future.  In addition to the \twco lines from $J=3-2$ to $J=6-5$, we also detect two lines each of \thco, CN, HNC, and \hcop, and a single line of HCN.  
While, as mentioned in \S\ref{stacking}, the composite spectrum cannot strictly be used to discern the strength of a faint line, all visually apparent lines are accounted for in Table~\ref{tab:lines}, and no obvious lines remain unidentified.  To extend the observed \twco SLED to low $J$, we include in Table~\ref{tab:lines} the average \twco(1--0) and \twco(2--1) luminosities of these sources as observed by the Australia Telescope Compact Array (ATCA) and reported in \citet{aravena13} and forthcoming papers.

Below, we analyze and discuss the detected lines in more detail.  Where applicable, we use the radiative transfer code \texttt{RADEX} \citep{vandertak07} to constrain the average ISM conditions in these galaxies.  \texttt{RADEX} iteratively solves for the line emission and level populations of a given molecule using the escape probability formalism, and provides results comparable to Large Velocity Gradient (LVG) codes \nocite{scoville74}.  All models use collisional rate coefficients from the \texttt{LAMDA} database \citep{schoier05} and are carried out for the geometry of an expanding sphere.  Since absolute line luminosities are proportional to the common redshift and value of \sopf chosen as a reference, we carry out our analysis using only line ratios, with the understanding that the line luminosities given in Table~\ref{tab:lines} should be taken as relative to a source with $\sopf = 15$\,mJy at $z=3$.  For typical dust continuum SEDs of the SPT sources, this corresponds to an apparent $\lir \sim 4 \times 10^{13}$\,\lsol, with \lir integrated from $8-1000$\,\um.  Typical magnifications range from $\mu = 5-20$ \citep{hezaveh13}.

The results of this line excitation modeling should be viewed with caution, given the varying number of sources which contribute to each line.  With a typical bandwidth of $\sim$130\,GHz in the rest frame, at most two lines each of \twco, \thco, HCN, HNC, CN, or \hcop can appear in the spectrum of any given source.  Thus, lines of the same molecule with $\Delta J_{\mathrm{up}} \geq 2$ can have no contributing sources in common.  Additionally, the averaging of spectra of many galaxies with varied gas conditions may distort the observed line ratios such that typical gas properties are unrecoverable.  However, we show in Appendix \ref{appstack} that conditions representative of the input conditions are recovered, subject to the degeneracies inherent in large velocity gradient modeling.  Differential magnification, in which different spectral lines are magnified by different amounts, could also skew the averaged line ratios.  However, a detailed accounting of the effects of differential magnification would require knowledge of the filling fractions of gas emitting in each transition and the lensing geometry of our sources, which is beyond the scope of this paper.  The quantities derived from the radiative transfer modeling are intended to be representative of the gas conditions which give rise to the observed lines.  Significant uncertainty in properties of the molecular gas will remain as a result of the limited sample size, the variation of fluxes between sources, and the limited number of species and transitions detected.  Gas conditions derived from \twco transitions in individual sources will be presented in a future publication.

\begin{deluxetable*}{lcccclccc} 
\tablecaption{Spectral Line Properties \label{tab:lines}} 
\startdata 
\tableline 
\\ 
Line & $\nu_\mathrm{rest}$ & $N_\mathrm{sources}$ & $L'$ & & Line & $\nu_\mathrm{rest}$ & $N_\mathrm{sources}$ & $L'$  \\ 
     & (GHz)                 &                      & ($10^9$ K km/s pc$^2$) & &         & (GHz)                 &                      & ($10^9$ K km/s pc$^2$) \\ 
\tableline 
\\ 
CO(1-0)$^a$ & 115.2712 & 5 & \textbf{296.6 $\boldsymbol{\pm}$ 16.5} & & CH 536 & 536.7614 & 10 & 9.6 $\pm$ 4.8 \\ 
CO(2-1)$^a$ & 230.5380 & 10 & \textbf{329.0 $\boldsymbol{\pm}$ 12.9} & & OH 425 & 425.0363 & 8 & -1.7 $\pm$ 2.9 \\ 
CO(3-2) & 345.7960 & 6 & \textbf{256.9 $\boldsymbol{\pm}$ 10.2} & & OH 446 & 446.2910 & 10 & 5.0 $\pm$ 2.8 \\ 
CO(4-3) & 461.0408 & 12 & \textbf{179.1 $\boldsymbol{\pm}$ 9.3} & & CN($N=3-2$) & 340.2478 & 5 & \textbf{11.4 $\boldsymbol{\pm}$ 3.5} \\ 
CO(5-4) & 576.2679 & 10 & \textbf{199.8 $\boldsymbol{\pm}$ 16.8} & & CN($N=4-3$) & 453.6067 & 10 & \textbf{9.1 $\boldsymbol{\pm}$ 2.9} \\ 
CO(6-5) & 691.4731 & 2 & \textbf{136.3 $\boldsymbol{\pm}$ 30.4} & & CN($N=5-4$) & 566.9470 & 11 & 6.0 $\pm$ 5.2 \\ 
$^{13}$CO(3-2) & 330.5880 & 5 & \textbf{12.8 $\boldsymbol{\pm}$ 3.6} & & CN($N=6-5$) & 680.2641 & 2 & -16.4 $\pm$ 6.0 \\ 
$^{13}$CO(4-3) & 440.7652 & 9 & \textbf{12.6 $\boldsymbol{\pm}$ 3.0} & & SiO(7-6) & 303.9270 & 4 & -2.6 $\pm$ 4.0 \\ 
$^{13}$CO(5-4) & 550.9263 & 9 & 2.8 $\pm$ 5.8 & & SiO(8-7) & 347.3306 & 6 & -5.1 $\pm$ 3.6 \\ 
$^{13}$CO(6-5) & 661.0673 & 4 & 6.3 $\pm$ 5.9 & & SiO(9-8) & 390.7284 & 7 & -4.9 $\pm$ 4.0 \\ 
C$^{18}$O(3-2) & 329.3305 & 5 & 1.4 $\pm$ 3.8 & & SiO(10-9) & 434.1196 & 8 & -0.0 $\pm$ 2.8 \\ 
C$^{18}$O(4-3) & 439.0888 & 7 & -6.0 $\pm$ 4.5 & & SiO(11-10) & 477.5031 & 11 & -3.4 $\pm$ 3.8 \\ 
C$^{18}$O(5-4) & 548.8310 & 9 & 6.1 $\pm$ 5.5 & & SiO(12-11) & 520.8782 & 11 & 8.8 $\pm$ 5.3 \\ 
C$^{18}$O(6-5) & 658.5533 & 4 & 11.8 $\pm$ 6.1 & & SiO(13-12) & 564.2440 & 11 & -7.7 $\pm$ 5.3 \\ 
$[$CI$]$(1-0) & 492.1606 & 13 & \textbf{47.5 $\boldsymbol{\pm}$ 3.7} & & SiO(14-13) & 607.5994 & 8 & 2.2 $\pm$ 6.6 \\ 
HCN(3-2) & 265.8864 & 2 & -14.2 $\pm$ 10.1 & & SiO(15-14) & 650.9436 & 4 & -10.8 $\pm$ 6.0 \\ 
HCN(4-3) & 354.5055 & 6 & \textbf{13.4 $\boldsymbol{\pm}$ 3.2} & & SiO(16-15) & 694.2754 & 2 & 9.8 $\pm$ 9.8 \\ 
HCN(5-4) & 443.1161 & 9 & 4.4 $\pm$ 2.8 & & SiO(17-16) & 737.5939 & 2 & -5.4 $\pm$ 9.5 \\ 
HCN(6-5) & 531.7164 & 10 & 3.9 $\pm$ 5.0 & & CS(6-5) & 293.9122 & 3 & 1.3 $\pm$ 6.0 \\ 
HCN(7-6) & 620.3040 & 7 & 7.2 $\pm$ 7.0 & & CS(7-6) & 342.8830 & 5 & 1.2 $\pm$ 3.4 \\ 
HCN(8-7) & 708.8770 & 2 & 25.7 $\pm$ 10.2 & & CS(8-7) & 391.8470 & 7 & 2.7 $\pm$ 3.7 \\ 
HNC(3-2) & 271.9811 & 3 & 2.2 $\pm$ 8.5 & & CS(10-9) & 489.7510 & 13 & -5.9 $\pm$ 3.5 \\ 
HNC(4-3) & 362.6303 & 5 & \textbf{16.2 $\boldsymbol{\pm}$ 4.2} & & CS(11-10) & 538.6888 & 10 & -2.4 $\pm$ 5.4 \\ 
HNC(5-4) & 453.2699 & 10 & \textbf{15.6 $\boldsymbol{\pm}$ 4.3} & & CS(12-11) & 587.6162 & 10 & 9.5 $\pm$ 5.4 \\ 
HNC(6-5) & 543.8976 & 10 & 12.8 $\pm$ 5.1 & & CS(13-12) & 636.5318 & 5 & 15.4 $\pm$ 9.0 \\ 
HNC(7-6) & 634.5108 & 5 & -4.4 $\pm$ 10.2 & & CS(14-13) & 685.4348 & 2 & -1.1 $\pm$ 7.6 \\ 
HNC(8-7) & 725.1073 & 2 & -3.2 $\pm$ 9.1 & & CS(15-14) & 734.3240 & 2 & 1.3 $\pm$ 9.3 \\ 
HCO$^+$(3-2) & 267.5576 & 2 & 8.5 $\pm$ 9.5 & & NH$_3$(1$_{0}$-0$_{0}$) & 572.4982 & 10 & 5.4 $\pm$ 5.9 \\ 
HCO$^+$(4-3) & 356.7342 & 6 & \textbf{11.7 $\boldsymbol{\pm}$ 3.2} & & N$_2$H$^+$(3-2) & 279.5117 & 2 & -22.2 $\pm$ 10.7 \\ 
HCO$^+$(5-4) & 445.9029 & 10 & 3.0 $\pm$ 2.8 & & N$_2$H$^+$(4-3) & 372.6725 & 5 & -1.8 $\pm$ 3.6 \\ 
HCO$^+$(6-5) & 535.0616 & 10 & \textbf{22.6 $\boldsymbol{\pm}$ 4.8} & & N$_2$H$^+$(5-4) & 465.8250 & 10 & 0.8 $\pm$ 3.3 \\ 
HCO$^+$(7-6) & 624.2085 & 7 & 16.0 $\pm$ 7.6 & & N$_2$H$^+$(6-5) & 558.9667 & 9 & 1.6 $\pm$ 5.2 \\ 
HCO$^+$(8-7) & 713.3414 & 2 & 7.7 $\pm$ 10.4 & & N$_2$H$^+$(7-6) & 652.0959 & 4 & -3.8 $\pm$ 6.0 \\ 
H$_2$O(5$_{1,5}$-4$_{2,2}$) & 325.1529 & 5 & -0.6 $\pm$ 3.8 & & N$_2$H$^+$(8-7) & 745.2103 & 2 & 9.0 $\pm$ 9.8 \\ 
H$_2$O(4$_{1,4}$-3$_{2,1}$) & 380.1974 & 6 & 7.0 $\pm$ 3.4 & & CCH(3-2) & 262.0042 & 1 & 38.1 $\pm$ 19.3 \\ 
H$_2$O(4$_{2,3}$-3$_{3,0}$) & 448.0011 & 10 & -2.0 $\pm$ 2.8 & & CCH(4-3) & 349.3387 & 6 & 7.2 $\pm$ 3.3 \\ 
H$_2$O(5$_{3,3}$-4$_{4,0}$) & 474.6891 & 12 & 8.2 $\pm$ 3.9 & & CCH(5-4) & 436.6604 & 7 & 5.0 $\pm$ 2.9 \\ 
H$_2$O(1$_{1,0}$-1$_{0,1}$) & 556.9360 & 9 & -15.5 $\pm$ 5.4 & & CCH(6-5) & 523.9704 & 11 & 13.9 $\pm$ 5.2 \\ 
H$_2$O(2$_{1,1}$-2$_{0,2}$) & 752.0331 & 2 & \textbf{31.4 $\boldsymbol{\pm}$ 8.8} & & CCH(7-6) & 611.2650 & 7 & 14.2 $\pm$ 6.7 \\ 
H$_2$O$^+$ 604 & 604.6786 & 8 & -5.7 $\pm$ 6.7 & & CCH(8-7) & 698.5416 & 2 & -4.0 $\pm$ 10.5 \\ 
H$_2$O$^+$ 607 & 607.2273 & 8 & 2.0 $\pm$ 6.5 & & H21$\alpha$ & 662.4042 & 4 & -4.4 $\pm$ 6.1 \\ 
H$_2$O$^+$ 631 & 631.7241 & 5 & 6.7 $\pm$ 8.4 & & H22$\alpha$ & 577.8964 & 10 & -0.8 $\pm$ 5.5 \\ 
H$_2$O$^+$ 634 & 634.2729 & 5 & 3.9 $\pm$ 8.6 & & H23$\alpha$ & 507.1755 & 11 & -6.1 $\pm$ 5.4 \\ 
H$_2$O$^+$ 721 & 721.9274 & 2 & 7.3 $\pm$ 9.2 & & H24$\alpha$ & 447.5403 & 10 & 0.8 $\pm$ 2.8 \\ 
H$_2$O$^+$ 742 & 742.1090 & 2 & \textbf{29.3 $\boldsymbol{\pm}$ 8.9} & & H25$\alpha$ & 396.9008 & 7 & -3.8 $\pm$ 3.5 \\ 
H$_2$O$^+$ 746 & 746.5417 & 2 & 21.8 $\pm$ 8.2 & & H26$\alpha$ & 353.6227 & 6 & -1.0 $\pm$ 3.2 \\ 
H$_2$O$^+$ 761 & 761.8188 & 2 & 6.9 $\pm$ 10.3 & & H27$\alpha$ & 316.4154 & 4 & -1.4 $\pm$ 3.8 \\ 
LiH(1-0) & 443.9529 & 9 & 0.6 $\pm$ 2.9 & & H28$\alpha$ & 284.2506 & 3 & -3.0 $\pm$ 7.7 \\ 
CH 532 & 532.7239 & 10 & 12.6 $\pm$ 4.9 & &  & & & \\ 
\enddata 
\tablecomments{All fluxes have been scaled to $z=3$ and \sopf = 15\,mJy, corresponding to $\lir \sim 5 \times 10^{13}$\,\lsol.  Lines with $S/N > 3$ are shown in bold.  For transitions with fine or hyperfine structure, only the main transition is listed, and the line may be referred to by frequency instead of quantum numbers for clarity.} 
\tablenotetext{a}{Derived from ATCA low-$J$ \twco observations} 
\end{deluxetable*}

\subsection{\twco and its Isotopologues}

\subsubsection{\twco SLED}
\begin{figure}[htb]%
\includegraphics[width=\columnwidth]{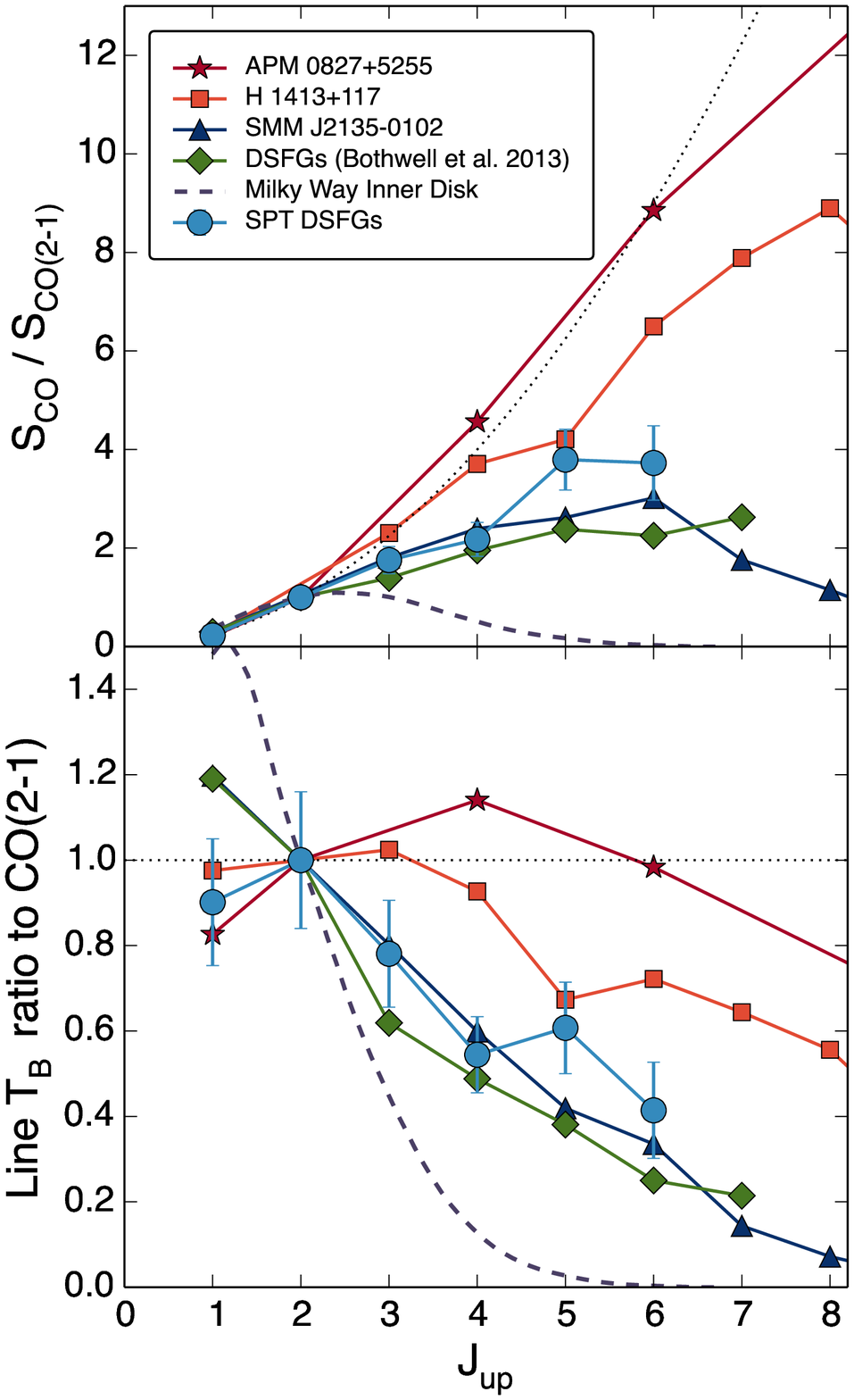}%
\caption{\twco SLEDs of the SPT DSFGs and other systems for comparison, normalized to the $J=2-1$ transition.  The average \twco(1--0) and (2--1) transitions are derived from data obtained with ATCA.  The SLEDs are plotted both in flux units (\textit{top}) and brightness temperature units (\textit{bottom}).  The SPT points are shown with an additional 15\% systematic uncertainty from the choice of weighting; uncertainties for the other objects are not shown.  The dotted line indicates thermalized, optically thick gas.  The two outlying objects, H1413+117 and \apm, both host powerful quasars.}%
\label{fig:cosled}%
\end{figure}

Figure \ref{fig:cosled} shows the composite \twco SLED in comparison to the well-sampled SLEDs of quasar H1413+117 ($z=2.56$, apparent $\lir = 2.4\times 10^{13}\lsol$; \citealt{barvainis94}), \smm ($z=2.32$, apparent $\lir = 3.8\times 10^{13}\lsol$; \citealt{swinbank10}), and quasar \apm  ($z=3.91$, apparent $\lir \sim 10^{15}\lsol$; \citealt{egami00}).  We also show the average SLED found by \citet{bothwell13} in a sample of (almost all unlensed) DSFGs selected at 850\um by SCUBA with optical or mid-IR spectroscopic redshifts.  The average SLEDs of both DSFG samples are clearly less excited than those of either high-$z$ QSO host, with the DSFG samples showing an apparent flattening or turnover near $J=5$ similar to local starburst galaxies M82 or NGC253 \citep{panuzzo10,bradford03}.  \citet{weiss07} and \citet{danielson11} use well-characterized (up to and beyond $J=10-9$) \twco SLEDs of \apm and \smm, respectively, to demonstrate multiphase interstellar media in those objects, while \citet{bradford09} find that the ISM of H1413+117 is well represented by a single, high-excitation gas component.  As our spectrum covers only six lines (or five line ratios) of \twco, none of which are clearly beyond the peak of the CO SLED, we postpone a discussion of excitation modeling in our objects until observations of \thco are added, below.

\subsubsection{\thco and \ceto}
Observations of \twco are frequently difficult to interpret because the high relative abundance of \twco molecules and low electric dipole moment cause the lines to become optically thick even at moderate densities ($\nHt \lesssim 10^3$\,\percc).  Observations of the less abundant isotopologues \thco and \ceto are thus useful, as these lines are frequently optically thin and the line strengths are then proportional to the total molecular column density.  For high-redshift objects, observations of \thco are also interesting due to the different formation mechanisms of the carbon and oxygen isotopes.  While \twc nuclei are produced during He burning in high-mass stars on rapid timescales, \thc nuclei are ``secondary,'' formed from \twc seed nuclei in intermediate-mass stars undergoing CNO cycle burning, at ages of $\gtrsim$1\,Gyr \citep{wilson94}.  The formation of \eo is less well understood, but the clearest route is also via a branch of the CNO cycle apparently most pronounced in massive stars, causing \eo-bearing species to be enhanced in galaxies with recent massive star formation \citep[e.g.,][]{henkel93}.  Identification of \thc-bearing species could be used as a type of nucleosynthesis chronometer, since the interstellar medium must be enriched by the metals from previous generations of stars for the species to be present.  At high redshift, then, the \twco / \thco line ratio could be large, since relativly little time will have passed for young starbursts to generate \thc nuclei and disperse them back into the interstellar medium \citep{hughes08,henkel10}.

In the stacked spectrum of SPT DSFGs, we detect both \thco(3--2) and \thco(4--3), but no transitions of \ceto.  Relative to their \twco counterparts, we find $L^{'}_{\twco} / L^{'}_{\thco} \sim 20$ and 15 in the $J=3-2$ and $J=4-3$ lines, respectively, indicating that the optical depth of \thco increases at least to $J=4-3$.  Our measurements of \ceto, meanwhile, are only sensitive enough to constrain $L^{'}_{\thco} / L^{'}_{\ceto} > 1$ ($3\sigma$) in the $J=3-2$ and $J=4-3$ transitions.

In Milky Way molecular clouds, typical \twco/\thco line ratios are $\sim$5--10 \citep[][e.g.,]{buckle10}, despite the variety of molecular cloud conditions from galactic center to the outer galaxy.  Through analysis of the optically thin wings of the CO lines, and in combination with observations of other molecules, the CO line ratio likely indicates a [\twc/\thc] abundance gradient from $\sim$25 near the galactic center to $\sim$100 beyond the Sun's galactocentric radius \citep{wilson94,wang09}.  In nearby spiral galaxies, the line ratio rises to $\sim$10 as typical large millimeter dish beams average over both molecular clouds and more diffuse regions where both CO isotopologues are optically thin.  Local infrared-luminous galaxies, on the other hand, show markedly weaker \thco lines -- \citet{greve09} find \twco/\thco line ratios of $\sim$40, 18, and 8 in the (1--0), (2--1), and (3--2) transitions, respectively, towards the prototypical ULIRG Arp~220, and weaker yet \thco lines in the LIRG NGC~6240.  These large line ratios have been interpreted as evidence of the merger-driven inflow of unenriched gas \citep{rupke08} or as a purely optical depth effect, in which the bulk of the \twco emission comes from a warm, turbulent medium with $\tau_{\twco} \lesssim 1$, while \thco arises from denser regions with a small filling factor \citep{aalto95}.  Typical \thco/\ceto ratios in the local universe generally range from $\sim 4-8$ \citep[e.g.,][]{henkel93}, which reflects the true abundance ratio of these species if both are optically thin.  This ratio decreases towards unity, however, in systems known to host recent star formation, with local ULIRG Arp~220 showing the lowest measured $\thco/\ceto = 1.0$ \citep{greve09}, which may indicate \eo enrichment from the massive stars in that system.

Few observations of CO isotopologues exist at high redshift.  \citet{danielson13} detect the (3--2) and (5--4) transitions of both \thco and \ceto in \smm, showing ratios to \twco of 18 and 43, respectively.  These authors, along with \citet{danielson11}, additionally constrain the (1--0), (4--3), and (7--6) transitions of \thco.  The most stringent of these leads to a 3$\sigma$ lower limit of \twco/\thco(4--3)$\gtrsim$60, a limit difficult to reconcile with the detection of other \thco lines at lower and higher $J$ in that object.  The \ceto transitions in \smm are unusually bright, comparable to the \thco lines as in Arp~220, which may indicate the presence of a large number of massive young stars.  \citet{henkel10} find \twco/\thco $\sim$40 in the $J=3-2$ line towards quasar H1413+117, suggesting a deficit of \thco on the order of [\twco/\thco] $= 300 - 10^4$.  If real, the lack of \thco would conflict with quasar absorption line studies in the optical and near-infrared, which have consistently shown solar or near-solar abundances of secondary species such as iron and nitrogen out to high redshift \citep[e.g.,][]{sameshima09,juarez09,nagao12}.  \citet{henkel10} posit that the apparent contradiction may originate from the geometry of the source, with the metal-enriched gas observed in the optical and infrared largely confined to the regions nearest to the AGN, while the CO emission arises from relatively metal-poor gas hundreds or more parsecs from the AGN.  The SPT DSFGs have \thco emission intermediate between that of local galaxies and H1413+117, similar to the line brightness ratios seen in \smm.  The SPT DSFGs do not, however, have \ceto emission as bright as in \smm, but further observations are necessary to determine the true strength of \ceto in the typical SPT DSFG.

To attempt to characterize the typical gas conditions and the [\twco/\thco] abundance ratio in the SPT DSFGs, we use \texttt{RADEX} to model the interstellar media of the SPT DSFGs in aggregate, under the assumption that all the detected \twco and \thco emission arises from gas of comparable temperature and density.  Since these models are compared not to the lines of a single object but rather to the average of lines of a large number of objects, we include an additional 15\% uncertainty in the calculated line ratios, motivated by the variation seen when different scaling methods are applied to the sources which contribute towards individual lines.  We run a grid of excitation models with free parameters $N_\mathrm{CO} / \Delta v$, the column density of \twco molecules per unit line width, $T_\mathrm{kin}$, the kinetic temperature of the gas, $n_\mathrm{H_2}$, the number density of molecular hydrogen, and [\twco/\thco], the relative abundance of the CO isotopologues.  Our grid of models spans $N_\mathrm{CO} / \Delta v = 10^{15}-10^{19}$\,\Ndv, $\tkin = 12-300$\,K, $\nHt = 10-10^6$\,\percc, and [\twco/\thco] $= 10-10^4$.  In all models, we fix the background temperature of the cosmic microwave background (CMB) to $T_\mathrm{CMB} = (1+z_\mathrm{common})T_\mathrm{CMB}(z=0) \sim 11$\,K, though, as we discuss further below, any value of $T_\mathrm{CMB}(z)$ provides nearly equivalent results for $z=2-6$.

Many prior assumptions typically implemented during line radiative transfer modeling -- for example, that the system be dynamically stable, or that the total gas mass not exceed the dynamical mass -- have no straightforward analog when modeling the average excitation of a collection of sources.  As such, we adopt only two priors.  First, motivated by the \texttt{RADEX} documentation, line optical depths are constrained to be $\tau < 100$, because at optical depths exceeding 100 the line excitation temperatures are unlikely to reflect conditions in the emitting region.  This prior effectively excludes the corner of parameter space defined by high \twco column density, low temperature, and low H$_2$ number density.  Second, we require that the gas have at least enough velocity dispersion to correspond to virialized motion under its own self gravity, i.e., that $K_\mathrm{vir}$ not be $\ll 1$.  A convenient parameterization of this criterion is provided in \citet{bradford09}.

\begin{figure*}[htb]%
\includegraphics[width=\textwidth]{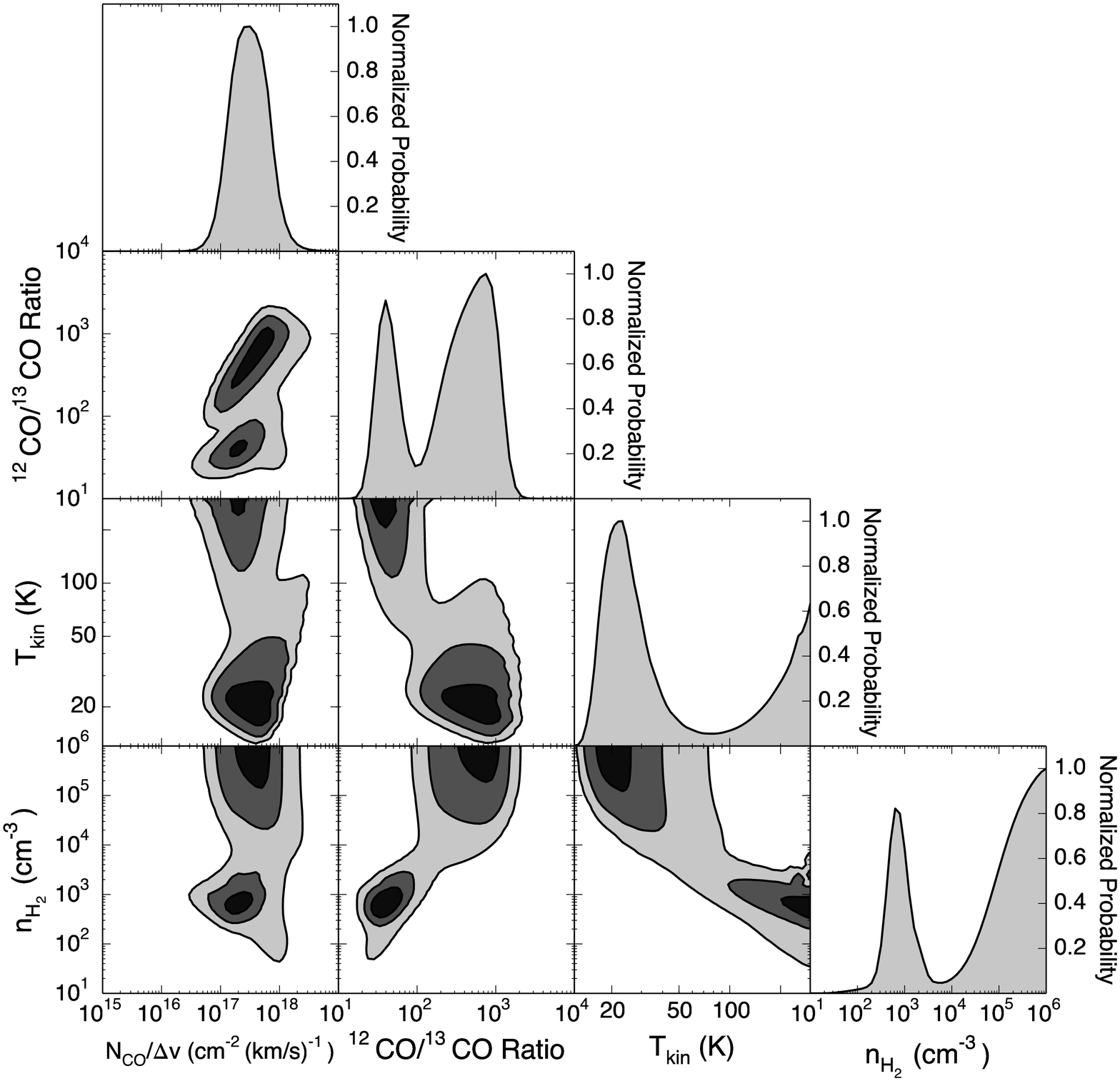}%
\caption{The results of our excitation analysis of the \twco and \thco lines detected in our composite spectrum, showing the degeneracies between parameters.  Marginalized posterior distributions for the parameters along the bottom axis are shown along the diagonal.  The \twco gas is thermalized at densities $\nHt > 10^6$\,\percc, and most \twco molecules are in $J>5$ levels for temperatures $\tkin > 150$\,K, so those regions of parameter space are equally well fit given the transitions available.  Contour levels are 1, 2, and 3$\sigma$.  The apparently unbounded one-dimensional marginalized parameter distributions are a direct consequence of degeneracies inherent in such modeling; expanded ranges would not lead to bounded distributions (see Appendix \ref{appstack}).}%
\label{fig:1213co}%
\end{figure*}

The parameter degeneracies and posterior probability distributions for each parameter are shown in Figure~\ref{fig:1213co}.  We find that no single model fits all seven available line ratios well (minimum $\chi^2 = 10.4$ for three degrees of freedom).  This likely reflects the simplicity of the modeling procedure, which represents a heterogeneous collection of galaxies as a single, homogeneous gas component of simple geometry.  The column density per velocity interval is well-determined, $\sim 3 \times 10^{17}$\,\Ndv, indicating that the \twco lines are moderately optically thick, with $\tau \sim 1 - 10$ for the transitions considered here.  The average line ratios as measured slightly prefer two solution ranges, with one solution implying a cold but very dense medium with $\tkin \sim 22$\,K, $\nHt > 10^{5}$\,\percc, and [\twco/\thco]$\sim 700$, while the other indicates a more diffuse but hot medium, with $\tkin \gtrsim 100$\,K, $\nHt \sim 800$\,\percc, and [\twco/\thco]$\sim 40$.  However, further investigation indicates that this apparent bimodality is largely due to the influence of the \twco(4-3) line.  If the flux of this line is increased by $\sim30\%$, the standard degeneracy between \nHt and \tkin is recovered.  A larger sample size and studies of individual DSFGs will indicate whether this apparent deficit of \twco(4-3) is in fact real or whether it indicates that the $J \geq 5$ \twco lines are beginning to trace a second, higher-excitation gas component.  Without further data, we cannot break the degeneracy between \tkin and \nHt.    If, however, the gas and dust are effectively coupled, the temperatures of gas and dust are roughly comparable.  In this case, with the measured range of dust temperatures in these sources $T_d = 20-50$\,K, we obtain a density of $\nHt \sim 10^4 - 10^5$\,\percc and an abundance ratio [\twco/\thco]$\sim 100-200$.  Measuring additional \twco lines would indicate whether the tentative flattening we see near $J=5$ indeed leads to a turnover at higher $J$, and may help break the degeneracies present in these radiative transfer models.  We note that such degeneracy is common and expected for CO SLEDs (and is illustrated clearly by \citealt{carilli13}, their Fig.~3, and Appendix \ref{appstack}).

With the above degeneracies in mind, for the typical submillimeter galaxy in the SPT sample, we find that \thco is likely no more than 3\% as abundant as \twco.  In the best fit solution range, the \thco lines have moderate opacity -- $\tau_{\thco} = 0.4$ for $\tkin = 40$\,K.  Better abundance constraints require observations of purely optically thin species, or observations of sufficient depth that the optically thin wings of a centrally optically thick line can be used for abundance measurements \citep[e.g.,][]{muller06}.

\begin{figure}[htb]
\includegraphics[width=\columnwidth]{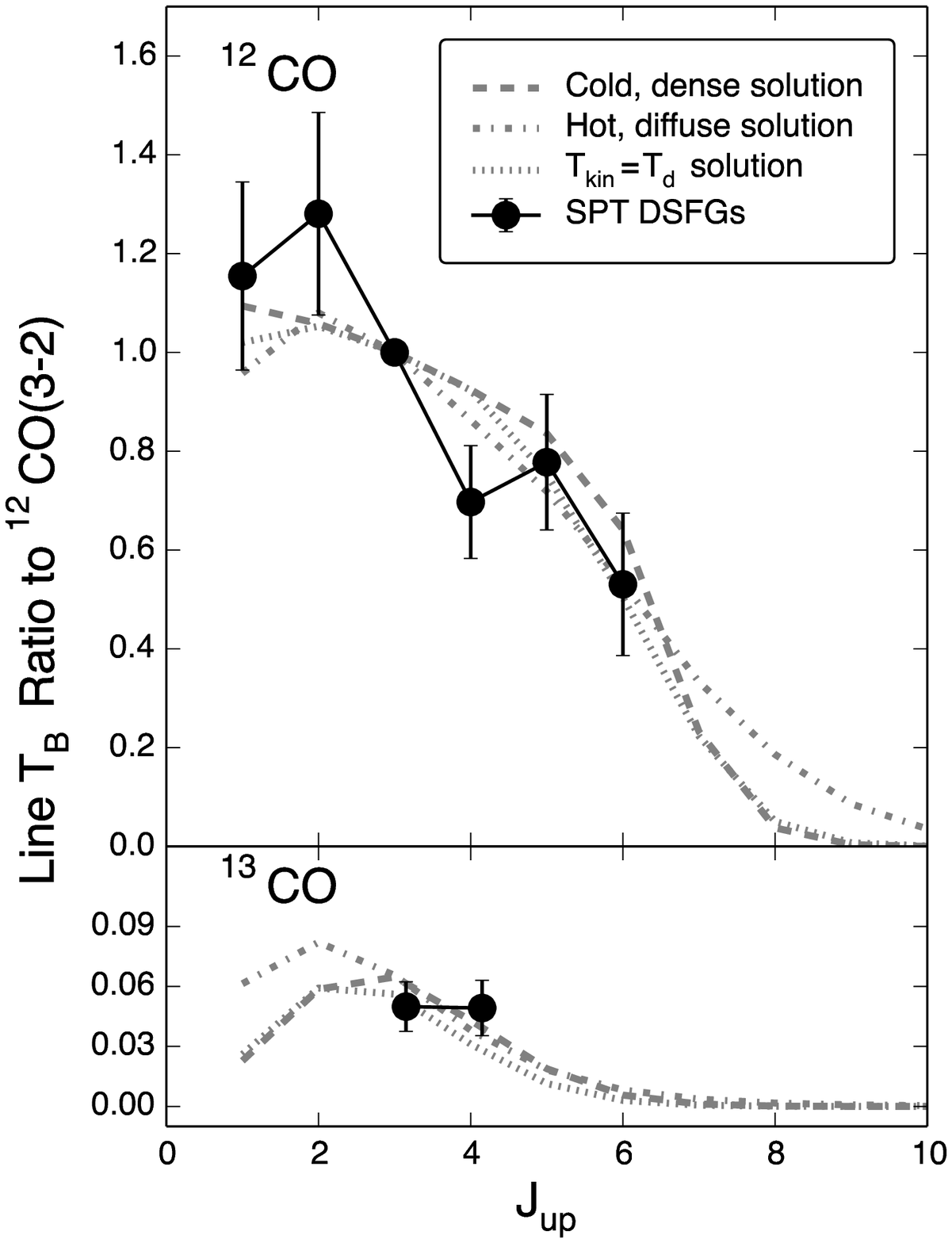}
\caption{Representative solutions to the excitation modeling of \twco and \thco marked as dashed lines,  from the parameter degeneracies shown in Figure~\ref{fig:1213co}.  The three solutions shown demonstrate the difficulty in distinguishing gas conditions along the degenerate curve in \nHt-\tkin.  Each panel shows the line ratio, in brightness temperature units, with respect to the \twco(3--2) line.  The top panels show the \twco SLED, the bottom panels the \thco SLED.}
\label{fig:sledsols}
\end{figure}

\subsection{The Dense Gas Phase}
Due to its weak electric dipole moment, \twco rapidly becomes thermalized at densities above a few hundred molecules\,\percc.  Molecules such as HCN, HNC, and \hcop, on the other hand, require densities $\sim 100 \times$ higher for collisional excitation, and have been used extensively within the Milky Way and in local galaxies as probes of the dense phase of the interstellar medium from which stars actively form.  In the rest-frame spectrum of SPT DSFGs, we detect 7 lines of CN, HCN, HNC, and \hcop from 240--780\,GHz.  Specifically, we detect the $(4-3)$ transition of all four species, the $N=3-2$ transition of CN, the $J=5-4$ HNC line, and the $J=6-5$ line of \hcop.  All lines are roughly equally bright, suggesting the emission from these lines is optically thick, nearly thermalized, and arises from a warm and dense medium.  Such thermalized emission would certainly be unusual were it to arise from a single object -- \citet{knudsen07} and \citet{papadopoulos07} both find that subthermal excitation of these dense gas tracers is common, with thermalized emission of HCN up to $J=4-3$ arising only in local ULIRG Arp~220.  

The relationship between $L^{'}_\mathrm{HCN}$(1--0) and \lir appears linear over a remarkable 7--8 orders of magnitude \citep{wu05}, from dense galactic cores to ULIRGs, suggesting that the same gas responsible for the HCN emission also gives rise to the infrared luminosity indicative of massive star formation.  This linear relation may break down for high-redshift objects \citep[e.g.,][]{gao07,riechers07b}, though many of the objects detected thus far also host active galactic nuclei which may provide an additional contribution to \lir at short wavelengths.  A similar relation has also been found for \hcop and the cyanide radical CN \citep{riechers06b,riechers07a}, though interpretation of CN is difficult due to its strong spin-spin and spin-nucleus coupling.  The fine and hyperfine structure transitions of CN lie close enough to each other ($\lesssim$few hundreds of km/s) as to be spectrally confused for most galaxy-integrated measurements, and we make no attempt to separate the components within our very wide channels.

In comparison with other molecular ISM tracers, $L^{'}_\mathrm{HCN}(1-0) / L^{'}_\mathrm{CO}(1-0)$ may indicate a type of ``dense gas fraction.''  With its low critical density, CO can be interpreted as tracing the total gas mass, while HCN, HNC, and \hcop trace only the densest active star-forming regions.  Support for this framework comes from observations showing that, in nearby spiral galaxies, $L^{'}_\mathrm{HCN} / L^{'}_\mathrm{CO} \sim 0.01 - 0.06$, rising to $\sim$0.15 in starbursting systems \citep{gao04b}.  \citet{juneau09} use observations of the $J=1-0$ and $J=3-2$ transitions of HCN and \hcop in local (U)LIRGs and numerical simulations to show that the dense gas fraction can rise sharply during the final stages of a major merger event.  In this scenario, the dissipation of angular momentum during the galactic collision funnels vast quantities of gas to the center of the system, triggering an intense burst of star formation.  In the average SPT DSFG spectrum, we find $L^{'}_\mathrm{HCN}(4-3) / L^{'}_\mathrm{CO}(1-0) = 0.045$.  For the range of HCN excitation found by \citet{papadopoulos07}, this corresponds to $L^{'}_\mathrm{HCN}(1-0) / L^{'}_\mathrm{CO}(1-0) = 0.045 - 0.165$, in agreement with typical measurements of actively star-forming galaxies \citep[][e.g.,]{gao07}.

The dense gas tracer lines we detect have comparable brightness, which would be unexpected if each molecule arose in the same gas conditions unless the lines are highly optically thick.  However, variations in line strength between the various dense gas tracers may be attributable to a variety of chemical abundance effects.  Models from \citet{maloney96}, for example, indicate that \hcop may be preferentially destroyed in X-ray Dominated Regions (XDRs), while more recent simulations from \citet{meijerink05} find instead that the abundance of HNC relative to HCN may increase in the presence of an XDR.  In these models, the ionization structure induced by the deeply-penetrating X-rays alters the balance of formation channels for the three molecules.  HNC may also become overluminous in the centers of galaxies as the dominant reaction type shifts from ion-neutral to neutral-neutral as density increases \citep{aalto02}.  The CN radical, meanwhile, which has a critical density intermediate between CO and \hcop, appears particularly well-suited to tracing regions with strong UV fields where HCN and HNC may be selectively dissociated \citep{boger05}, as in the filaments of Orion A \citep{rodriguezfranco98}.  As a further complication, the high-$J$ lines of HNC, HCN, and \hcop may not result from purely collisional excitation, as the presence of a strong mid-IR radiation field allows these molecules to be radiatively pumped via their first bending modes at 21, 14, and 12\,\um, respectively.  We discuss this possibility in more detail in \S\ref{highex}.

At high redshift, the faintness of these molecular lines has historically confined their study to bright, highly magnified lensed systems, with only two objects detected in multiple transitions or species.  In quasar H1413+117, observations of low- and mid-$J$ transitions of HCN, \hcop, and CN indicate purely collisional excitation with $\tkin \sim 60$\,K and $\nHt \sim 10^{4.8}$\,\percc \citep{wilner95,barvainis97,solomon03,riechers06b,riechers07a,riechers11d}.  The \apm quasar, in contrast, is bright in HCN, HNC, and \hcop at least to the $J=6-5$ lines, and the excitation of those species indicates that radiative pumping is likely in play \citep{garciaburillo06b,guelin07,weiss07,riechers10b}.  Such a finding is not unexpected, given the very hot dust ($T_d \sim 220$\,K) surrounding the nuclear starburst / active nucleus in this system.

To investigate the gas conditions which could excite these lines, we again make use of \texttt{RADEX} models.  We assume the emission from all four of these species comes from the same single gas component, not necessarily related to the origin of the CO emission discussed previously.  While the medium that gives rise to these dense gas tracer lines will also contribute to the CO emission, the converse is less likely to be true, given the excitation requirements of higher critical density molecules.  In order to narrow the vast parameter space that emerges when many species are modeled simultaneously, we assume the relative abundance ratios of CN, HCN, HNC, and \hcop are the same as in the local starburst galaxy NGC253 -- namely, [\hcop/HCN] = 2, [HNC/HCN] = 0.5, and [CN/HCN] = 0.25 \citep{wang04}.  Since we only aim to infer general properties for DSFGs, this choice is not critical; using abundance ratios as derived for the prototypical starburst galaxy M82 \citep{naylor10}, or even equal abundances of all four species changes the results by less than an order of magnitude in column and number densities and a factor of two in gas kinetic temperature.  As before, we implement a prior on the line optical depth $\tau$, require solutions with enough velocity dispersion to correspond to virialized motion under the self-gravity of the gas, and include an additional 15\% uncertainty in the modeled line ratios to account for the varied sources contributing to the flux of each line.

\begin{figure}[htb]%
\includegraphics[width=\columnwidth]{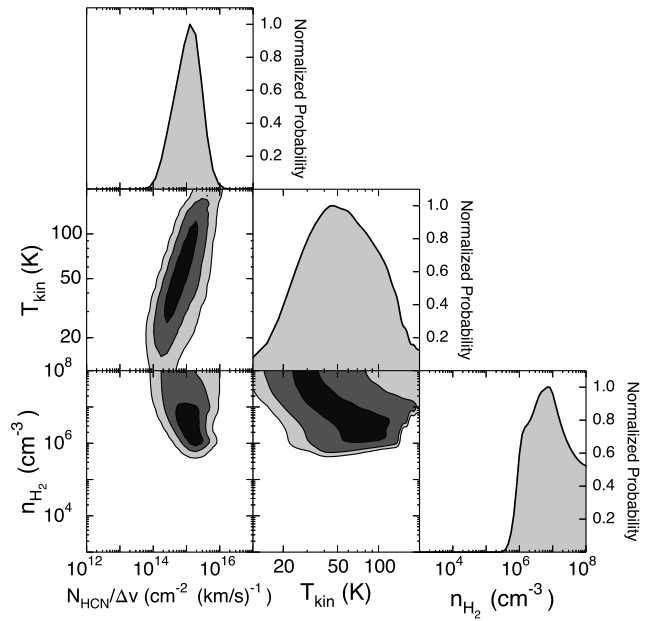}%
\caption{Parameter degeneracies in the excitation analysis of HCN, HNC, \hcop, and CN, as in Fig.~\ref{fig:1213co}.  Contour levels are 1, 2, and 3$\sigma$.  Again, the unbounded one-dimensional marginalized distributions reflect inherent degeneracies in the radiative transfer models, explored further in Appendix \ref{appstack}.} %
\label{fig:dense}%
\end{figure}

Our analysis indicates that the emission from mid-$J$ lines of HCN, HNC, \hcop, and CN arises largely from warm, dense gas, with $\tkin \sim 55$\,K and $\nHt \sim 10^{5.5}-10^{7.5}$\,\percc (Fig.~\ref{fig:dense}), though significant degeneracies remain.  The best-fit molecular column density per velocity interval, $N_\mathrm{HCN} / \Delta v \sim 10^{15}$\,\Ndv, is high enough to indicate that the lines of species other than CN are moderately optically thick, with $\tau \sim 5-30$ (due to its many fine and hyperfine energy level splittings, CN is always optically thin).  The high optical depth reduces the critical density for collisional excitation by a comparable factor.  The prior constraint that $\tau$ not be too high effectively constrains the molecular hydrogen density to $\nHt > 10^5$\,\percc.  The extreme optical depths which would be required to fit the observed line ratios for $\nHt < 10^5$\,\percc lead us to investigate alternatives to pure collisional excitation in \S\ref{highex}, below.

As a cross-check, we can calculate the fraction of the total \twco emission which arises from this dense gas component by assuming the \twco lines emitted by gas in the dense phase have the same filling factor as the lines of the dense gas tracer molecules.  The assumption of equal filling factors effectively provides a normalization between the dense gas tracer lines and the \twco lines.  Assuming a [\twco/HCN] abundance ratio of 8000 as in NGC253 \citep{wang04}, the estimated contribution of \twco emission arising from this gas component to the total is $4-10\%$ from \twco(1-0) to \twco(6-5).  This confirms that fitting the two components separately is justified.  The gas traced by these high-critical density molecules only begins to dominate the \twco emission beyond the $J=8-7$ transition; future radiative transfer modeling of individual DSFGs will require multiple gas components when such high-$J$ \twco lines are included.

\section{Discussion} \label{discussion}

\subsection{Comparison to other High-Redshift Galaxies}
Given the substantial observing time invested in the pre-ALMA era to detect the faint ISM diagnostics presented here, it is instructive to place our results in context with other well-studied high-redshift objects.  We focus on three systems in particular: the lensed starburst galaxy \smm, 
and lensed quasars H1413+117 and \apm.  
These objects may not be representative of the high-redshift star-forming galaxy population.  \apm, in particular, appears to be one of the most extreme objects in the universe by any measure.  Our results, based on the aggregate properties of many sources, should be relatively immune to distortion by extreme outlying objects, allowing the average properties of DSFGs to be determined.

\begin{figure*}[htb]%
\includegraphics[width=\textwidth]{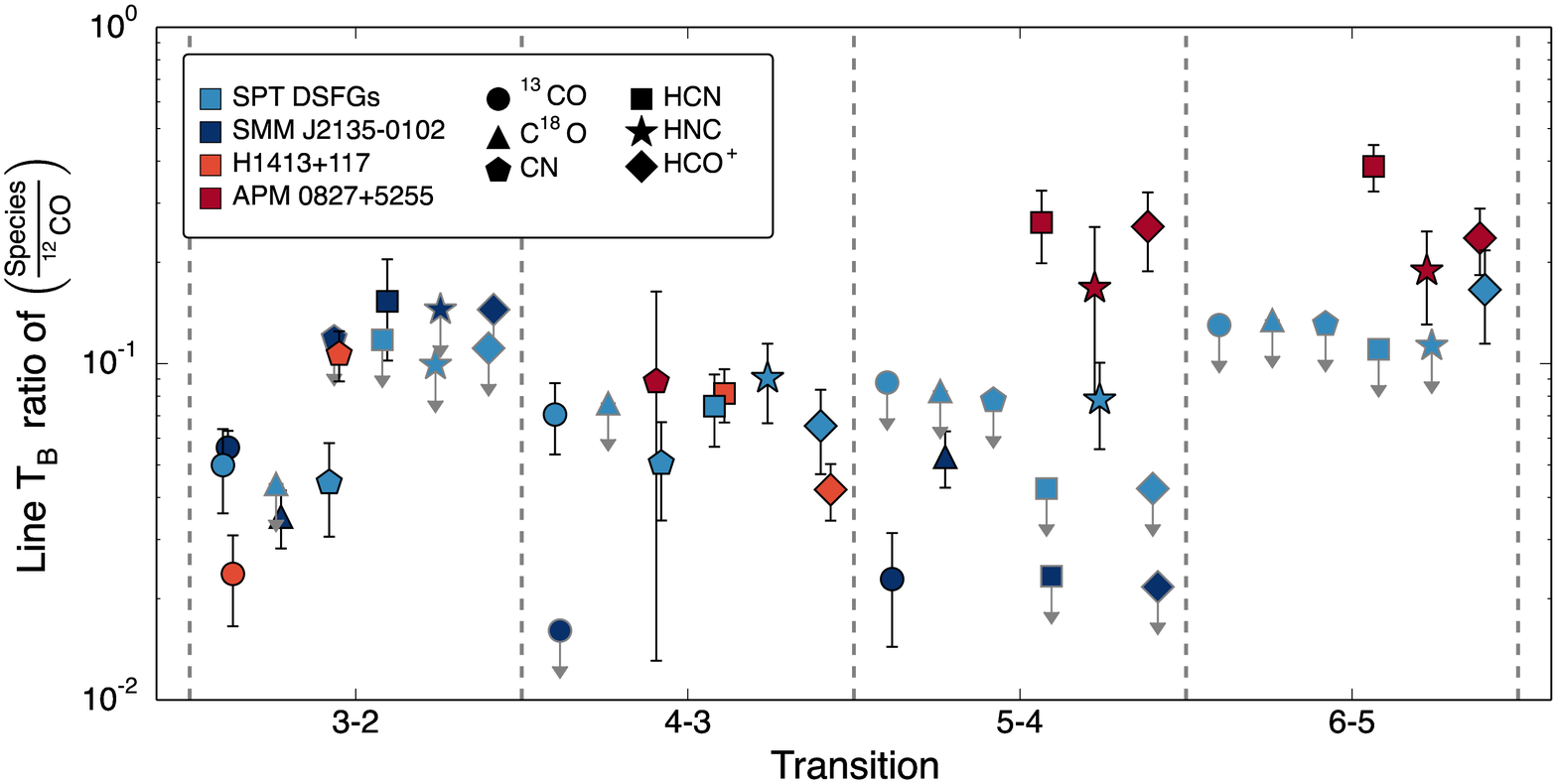}%
\caption{Comparison of faint lines in the SPT DSFGs to those detected in three high-redshift lensed sources: starburst galaxy \smm, and quasars  H1413+117 and \apm.  Plotted is the ratio (in brightness temperature units) of the faint line to the same-$J$ \twco line -- for example, \thco(4--3)/\twco(4--3) or CN($N=3-2$)/\twco($J=3-2$).  All upper limits are $3\sigma$.  Data for \smm from \citet{danielson11,danielson13}; for H1413+117 from \citet{barvainis97,weiss03,riechers07a,bradford09,henkel10,riechers11d}; for \apm from \citet{downes99,wagg05,garciaburillo06b,guelin07,weiss07,riechers10b}.  The \twco(5-4) line luminosity for \apm is interpolated from the CO SLED presented in \citet{weiss07}.}%
\label{fig:comparison}%
\end{figure*}

ALMA will dramatically reduce the time necessary to detect molecular ISM lines at high redshift -- summed across sources, all lines we detect in our average spectrum were observed for less than 30 minutes in total, and with only $\sim$1/4 the final number of antennas incorporated into the array.  Given that each of our sources was observed for only two minutes per tuning, and scaling to the noise level reached for the stacked transitions, we find that at full capacity, ALMA will be able to detect these faint lines and at higher spectral resolution in only half an hour per line, assuming a fiducial target galaxy with $\lir = 5 \times 10^{13}$\,\lsol.  As a tool for planning future ALMA observations of high-redshift DSFGs, we show in Fig.~\ref{fig:comparison} the ratio of the line luminosity of several species to the \twco line luminosity at the same rotational level for our composite spectrum and the reference objects described above.  The SPT DSFGs exhibit brighter \thco emission than H1413+117, but are fainter in most other lines which trace the dense phase of the ISM.  Both the SPT DSFGs and \smm appear to have significantly less-highly excited high-critical density molecules at $J>4$, underscoring the extreme conditions present in the circumnuclear region of the \apm quasar.

\subsection{Combining Multiple Transitions of Each Species}
For many species, we detect no single transition significantly, but by stacking all transitions of those species that fall within our bandpass, we can constrain the total luminosity emitted by a given molecule.  As an example, there are 10 hydrogen-$\alpha$ radio recombination lines (H20$\alpha$ to H29$\alpha$) from 250--770\,GHz.  In Table~\ref{tab:stackedlines}, we give the average integrated flux per transition of a number of species which have numerous lines within our bandpass.  Such a table is intended to draw attention to species which are ripe for future study.  

Aside from the lines with individually detected transitions, we also find significant flux in the stacked transitions of the rigid rotor CCH and in the CH $N,J = 1,3/2 \rightarrow 1,1/2$ doublet at 532 and 536\,GHz, the first detection of these species at high redshift.  The formation mechanisms of both molecules -- for CH, the radiative association of C$^+$ and H$_2$; for CCH, the dissociative recombination of C$_2$H$_2^+$ or C$_2$H$_3^+$ -- indicates that they likely trace regions of moderate density ($n \sim 10^3$\,\percc) where UV radiation controls the chemical networks \citep{sheffer08,godard09,gerin10,gerin11}.  Given the close proximity of the CH doublet lines to the $J=6-5$ lines of HCN and \hcop, all three species can be observed simultaneously in the 3\,mm atmospheric window for $z\sim4$ sources, making this group of transitions a potentially powerful probe of the chemistry of the ISM at high redshift.

Using Table~\ref{tab:stackedlines}, we can quantify the amount of contamination to brighter lines from molecules with closely spaced frequency ladders.  As an example, based on unexpectedly luminous \hctn emission up to $J=25-24$ in the highly-obscured LIRG NGC~4418 \citep{aalto07}, \citet{riechers10b} speculate that the \hcop(5--4) flux in the \apm quasar may be contaminated by emission from \hctn(49--48).  From 40 transitions of \hctn well-separated from other lines, observed nearly 250 times, we find no evidence for significant emission from \hctn, indicating that this heavy rotor does not significantly contaminate the flux of other observed lines at high-$J$.

\begin{deluxetable}{lccc} 
\tablecaption{Average Combined Line Properties \label{tab:stackedlines}} 
\startdata 
\tableline 
\\ 
Species & Transitions & Observations & $\left\langle L^{'} \right\rangle$ \\ 
 & & & ($10^9$ K km/s pc$^2$) \\ 
\tableline 
\\ 
$^{12}$CO & 4 & 30 & \textbf{209.3 $\boldsymbol{\pm}$ 6.2} \\ 
$^{13}$CO & 4 & 27 & \textbf{10.6 $\boldsymbol{\pm}$ 2.0} \\ 
C$^{18}$O & 4 & 26 & 1.2 $\pm$ 2.3 \\ 
HCN & 6 & 36 & \textbf{7.5 $\boldsymbol{\pm}$ 1.8} \\ 
HNC & 6 & 35 & \textbf{12.1 $\boldsymbol{\pm}$ 2.3} \\ 
HCO$^+$ & 6 & 37 & \textbf{9.8 $\boldsymbol{\pm}$ 1.8} \\ 
CS & 4 & 57 & 0.8 $\pm$ 1.6 \\ 
H$_2$O$^+$ & 7 & 34 & 7.5 $\pm$ 2.9 \\ 
CH & 2 & 20 & \textbf{11.3 $\boldsymbol{\pm}$ 3.4} \\ 
OH & 2 & 18 & 1.8 $\pm$ 2.0 \\ 
CN & 4 & 28 & \textbf{9.0 $\boldsymbol{\pm}$ 2.1} \\ 
SiO & 10 & 75 & -2.4 $\pm$ 1.4 \\ 
N$_2$H$^+$ & 6 & 34 & -0.7 $\pm$ 2.0 \\ 
H$\alpha$ RRLs & 10 & 56 & -1.8 $\pm$ 1.4 \\ 
CCH & 6 & 34 & \textbf{7.8 $\boldsymbol{\pm}$ 1.9} \\ 
H$_2$CO & 9 & 56 & 1.8 $\pm$ 1.3 \\ 
HC$_3$N & 40 & 249 & -1.5 $\pm$ 0.8 \\ 
\enddata 
\tablecomments{Luminosity given is the \textit{average} luminosity per transition; luminosities vary for individual transitions.  Line averages exclude individual transitions which are strongly blended with other lines.  The number of observations refers to the number of times a line was observed in a single source, summed over all sources and transitions for each species.  Molecules with combined transitions detected at $S/N > 3$ are shown in bold.}
\end{deluxetable}

\subsection{Excitation in Dense Gas Tracer Molecules}\label{highex}
Oddities exist in the line ratios of the dense gas tracer species we have detected -- the HNC(4--3) and (5--4) lines are nearly equally luminous, and $L^{'}_{\hcop(6-5)}$ significantly exceeds $L^{'}_{\hcop(4-3)}$.  Given the high critical densities of these lines ($\ncrit > 10^7$\,\percc) which argue against purely collisional thermalization, alternate excitation mechanisms and geometrical effects must be explored.

The first possible mechanism, which is also the easiest to rule out, is that the changing CMB temperature affects the level populations of each observed line differently or that diminishing contrast between the lines and CMB alters the observed line ratios \citep{dacunha13}.  Since our redshift search program covered a fixed observed frequency range, the highest-$J$ molecular lines come preferentially from the highest-redshift sources, in which the higher CMB temperature can more easily populate the upper energy states.  This is unlikely to offer a satisfactory explanation for our observed line ratios.  From our lowest-redshift object to the highest, the CMB changes temperature only by a factor of two, from 8.2--18\,K, and this temperature is still much lower than other relevant temperatures in these systems, for which $\tkin \sim T_d \gtrsim 30$\,K.  Additionally, if the dust and gas are nearly in thermodynamic equilibrium, for a given dust temperature at $z=0$, the thermodynamically equivalent temperature at higher redshift is $(\tkin^6(z=0) + T_\mathrm{CMB}^6(z))^{1/6} \approx \tkin(z=0)$ so long as $T_d = \tkin > T_\mathrm{CMB}$, assuming a dust emissivity law of $\beta = 2$ \citep{papadopoulos00}.

Of course, with a sample size of less than 25 objects, the ``average'' properties we derive may still be dominated by outliers.  The faint lines we detect typically arise from spectral regions covered by 5--10 sources, and using a weighted average increases the contribution from bright sources.  In the least-sampled lines, a single source contributes $\sim65\%$ of the weight to the stacked line, though out of just five or six sources.  Whether our observed line ratios are truly typical of the DSFG population (lensed and/or unlensed) will become clearer as other targeted spectroscopic observations and blind redshift surveys are performed.

Alternatively, the excitation of HNC, HCN, and \hcop may be strongly affected by pumping by mid-infrared photons.  In this scenario, a molecule in the ground vibrational state, $\nu=0$, $J=J_i$, absorbs a mid-IR photon to excite it to the $\nu=1$, $J=J_i+1$ state before decaying to the $\nu=0$, $J=J_i + 2$ state, a net change of $\Delta J = +2$.  The first bending modes of HNC, HCN, and \hcop are at 21, 14, and 12\um, respectively, so for dust-dominated continuum SEDs, which fall from 21--12\um, we might expect HNC to be more susceptible to mid-IR pumping than \hcop.  Various studies of local IR-bright galaxies have concluded that this effect is unimportant compared to collisional excitation \citep{stutzki88,gao04b}, but the phenomenon is still plausible for the \apm quasar, where very hot $T_d \sim 220$\,K dust near the central AGN creates an intense radiation field in the mid-IR \citep{riechers10b}.  Whether this effect is also possible in the SPT DSFGs is less clear.  No SPT source in the sample presented here shows a continuum SED with $T_d > 60$\,K.  Detection of mid-IR absorption lines would provide definitive proof of the effectiveness of mid-IR pumping \citep[e.g.,][]{lahuis07}, but such observations would be extremely challenging given the faintness of these sources.  Instead, we use \textit{Herschel}/PACS measurements at 100 and 160\um to constrain the short wavelength side of the dust SEDs (Strandet et~al., in prep.), which we accomplish by allowing $\lambda_0$, the wavelength at which the dust opacity is unity, to be a free parameter (which had previously been fixed at 100\,\um).  Since we aim only to divide the sample into two groups, the precise form of this fitting function is unimportant; nearly the same source divisions emerge when the short-wavelength side of the continuum SED is represented by a power law with index $\alpha=2$.

\begin{figure}[htb]%
\includegraphics[width=\columnwidth]{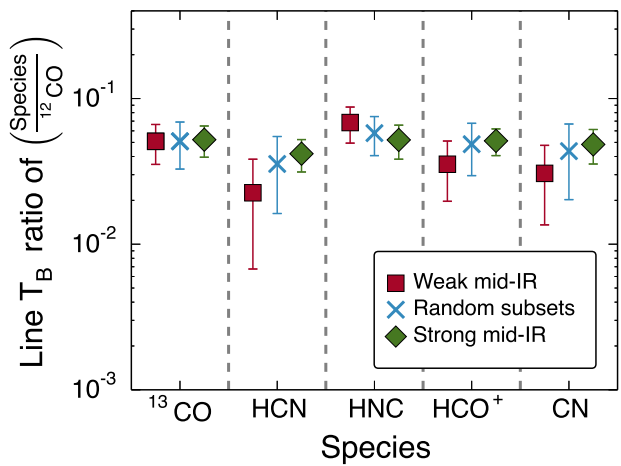}%
\caption{Line ratios in comparison to \twco determined by dividing our sample into a group with strong mid-IR flux density ($S_{21\um} > 2$\,mJy) and a group with weak mid-IR flux density ($S_{21\um} < 2$\,mJy).  Also shown is the range of values (average and 16-84\% interval) found when randomly drawing subsamples of half the sources (with replacement). That the high critical density molecules HNC, HCN, and \hcop are comparably bright in the two groups indicates that these molecules are unlikely to be pumped to high-$J$ energy levels via their first vibrational modes in the mid-IR.  Note that HCN, \hcop and CN are detected at only 2.0, 2.0, and 2.5$\sigma$ in the low mid-IR group of sources, respectively.} %
\label{fig:irpump}%
\end{figure}

With these short-wavelength SED fits in hand, we again scale all SEDs to the common redshift of $z=3$ before dividing the sample based on rest-frame 21\um flux density.  For our sample, we divide the sources into two groups, with $S_{21\um} > 2$\,mJy and $S_{21\um} < 2$\,mJy.  Both groups of sources span our entire redshift range, and the grouping is not influenced by PACS detection or non-detection (that is, sources with PACS 100\um detections are present in both groups, as are those with non-detections).  Both H1413+117 and \apm would fall in the high mid-IR group of sources (although the continuum SEDs of our sources are less well constrained in the rest-frame mid-IR than either of these objects), and \apm has a stronger mid-IR radiation field than any of the SPT sources.  Since dividing the sample in half lowers the number of sources for each individual transition to 2--5, we average all available transitions of each molecule to lower the uncertainty on the average line flux.  To determine the significance of any difference between the high and low mid-IR groups, we also repeat the above procedure for 1000 random subsets of the sample, each containing half the total number of sources.  The spread of this distribution shows the expected variation in line ratios when the sample is divided in half at random.  As shown in Fig.~\ref{fig:irpump}, both groups are consistent with randomly dividing the sample in half, suggesting that vibrational mode pumping is insignificant in the average SPT DSFG.

Finally, the SPT DSFGs could be subject to differential magnification which distorts the observed line fluxes.  Although lensing itself is wavelength-independent, the finite size of the background source causes different regions of the lensed object to be magnified by varying amounts, leading to variations in the ratios of observed spectral lines \citep[e.g.,][]{downes95}.  
This is an important effect when the tracers in question fill different parts of the target galaxy \citep{serjeant12,hezaveh12a}. However, for these chemically related and energetically similar species, differential magnification seems unlikely to influence the line ratios we have measured.

\section{Conclusions}
We have presented the average rest-frame millimeter spectrum of 22 gravitationally lensed dusty, star-forming galaxies at high redshift.  Stacking wide-band ALMA spectra of objects from $z=2-5.7$ allows faint ISM diagnostics to be studied as typical characteristics of the SPT DSFGs.  Aside from bright transitions of \twco, we also find multiple much fainter molecular transitions from \thco, HCN, HNC, \hcop, and CN.  The \twco SLED resembles that of other high-redshift DSFGs, and in conjunction with multiple \thco lines, we constrain the [\twco/\thco] abundance ratio to be $\sim 100-200$ for gas temperatures comparable to the dust temperatures of these sources, $\tkin \sim 40$\,K.  The average SPT DSFG has \thco brightness comparable to the only other high-$z$ star-forming system in which it has been measured (\smm), but does not show similarly bright \ceto.  Our excitation analysis indicates the HNC, HCN, \hcop, and CN emission arises from a warm, dense, and optically thick medium, which allows the energy levels 40--90\,K above the ground state to be sufficiently populated.  Population of vibrational modes in these molecules is unlikely to be relevant for the typical SPT DSFG, though a small sample size and the possibility of differential magnification may explain the line ratios we observe.  These observations represent the first constraints on the relative strengths of these lines in high-redshift star-forming galaxies, and will be instrumental in planning future ALMA observations of such systems.

\acknowledgements
{
The SPT is supported by the National Science Foundation through grant ANT-0638937, with partial support through PHY-1125897, the Kavli Foundation and the Gordon and Betty Moore Foundation. This paper makes use of the following ALMA data: ADS/JAO.ALMA \#2011.0.00957.S and \#2011.0.00958.S.  ALMA is a partnership of ESO (representing its member states), NSF (USA) and NINS (Japan), together with NRC (Canada) and NSC and ASIAA (Taiwan), in cooperation with the Republic of Chile. The Joint ALMA Observatory is operated by ESO, AUI/NRAO and NAOJ.  The National Radio Astronomy Observatory is a facility of the National Science Foundation operated under cooperative agreement by Associated Universities, Inc. The Australia Telescope Compact Array is part of the Australia Telescope National Facility which is funded by the Commonwealth of Australia for operation as a National Facility managed by CSIRO.  This research has made use of NASA's Astrophysics Data System.

}

\bibliographystyle{apj}
\bibliography{../bibtex/spt_smg}

\clearpage

\appendix
\section{Redshift Confirmation of SPT~0125-50}\label{appendix}

\begin{figure}[htb]%
\center{\includegraphics[width=0.6\columnwidth]{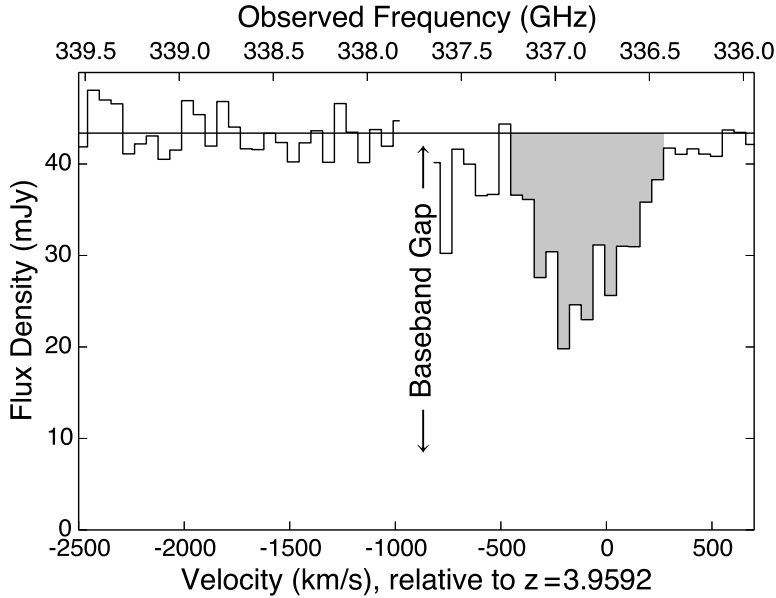}}
\caption{ALMA 870\,\um spectrum of SPT~0125-50 showing the serendipitous detection of the \hto($2_\mathrm{1,2} - 1_\mathrm{0,1}$) line in absorption, which confirms the redshift as $z=3.9592$.  The continuum level derived from the line-free basebands is shown as a horizontal line, at $\sim$44\,mJy.} %
\label{fig:spt0125-50}%
\end{figure}

The redshift of SPT~0125-50 was ambiguous in the ALMA 3\,mm spectra obtained during Cycle~0, as only the \twco(4--3) line was clearly detected ([CI] was only tentatively detected).  Its redshift has been confirmed as $z=3.9592$ by the serendipitous detection of the ortho-\hto($2_\mathrm{1,2} - 1_\mathrm{0,1}$) line, $\nu_\mathrm{rest} = 1669.90477$\,GHz, in the high-resolution 870\,\um ALMA Cycle~0 dataset presented in \citet{vieira13}.  The line is seen in absorption, as is common in local systems \citep[e.g.,][]{gonzalezalfonso12}.  The detection is shown in Figure~\ref{fig:spt0125-50}.

\section{Radiative Transfer Modeling of an Inhomogeneous Collection of Objects}\label{appstack}

One concern when performing radiative transfer modeling on an ensemble of sources is that variations in the ISM conditions between sources will render any inference about conditions from the averaged SLED meaningless.  An analagous problem applies to galaxy-integrated measurements, at least when fitting with a single gas component, as individual galaxies are also comprised of molecular clouds with a range of temperatures, densities, etc.  In the case of the SPT DSFGs, the present sample may contain objects with a range of CO excitation, from relatively quiescent gas conditions like those present in the Milky Way to the extreme excitation seen in some high-redshift quasars.  We stress that the radiative transfer embodied by the \texttt{RADEX} modeling is intended to determine only approximate conditions that would match the limited observations given a simplified geometrical model of the source.  Such simplifications apply in nearly all extragalactic and many galactic applications of the large velocity gradient radiative transfer approximation (of which \texttt{RADEX} is one open-source incarnation), which has now been widely used since Scoville \& Solomon (1974).  In this Appendix, we show that the \texttt{RADEX} modeling performed on a collection of sources recovers gas conditions that are representative of the input sources, subject to the degeneracy inherent in large velocity gradient models.  Here, we consider only \twco for simplicity, but the conclusions are easily extended to the other molecular tracers.

\begin{figure}[htb]
\centering
\includegraphics[width=0.49\textwidth]{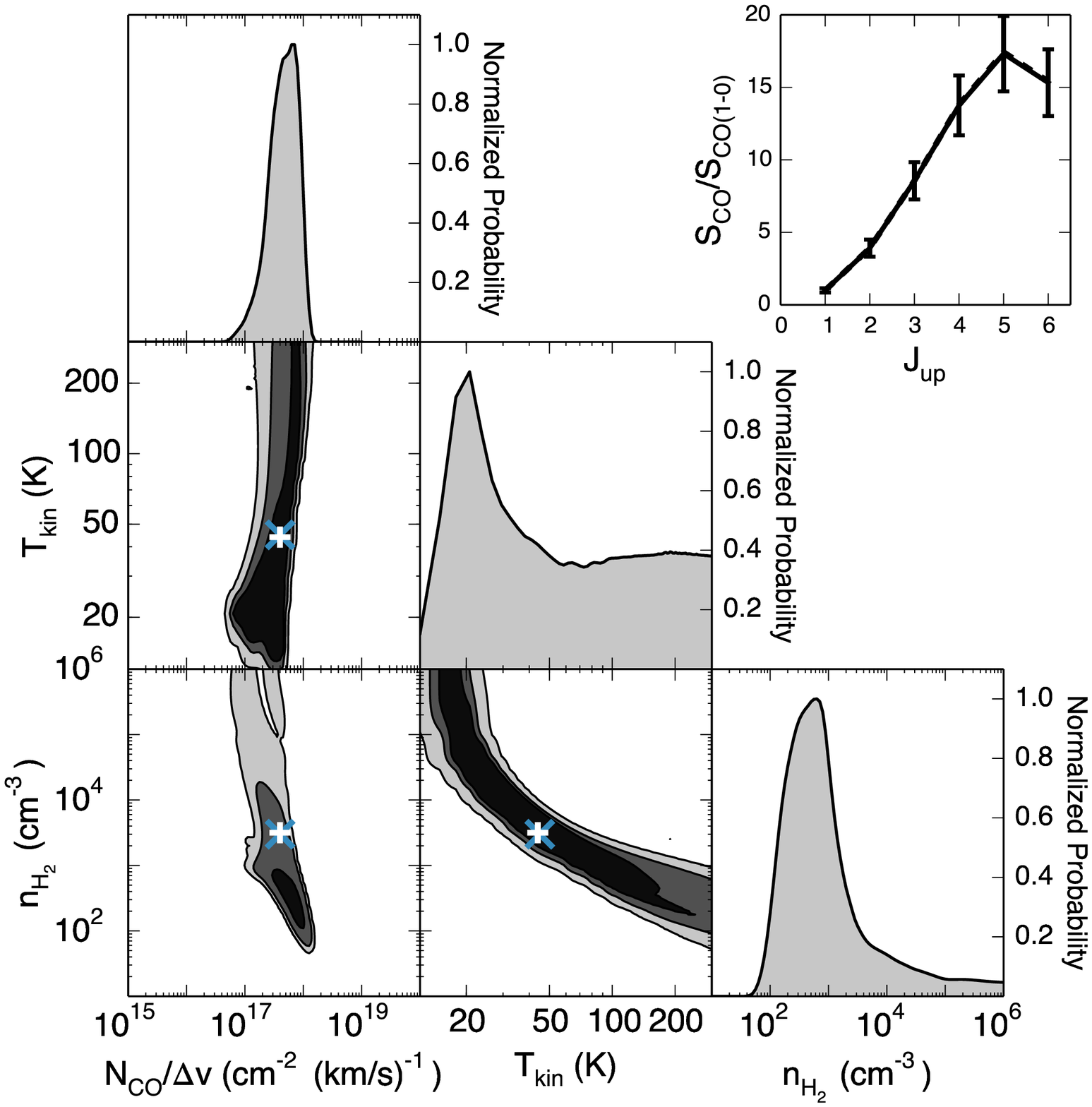}
\includegraphics[width=0.49\textwidth]{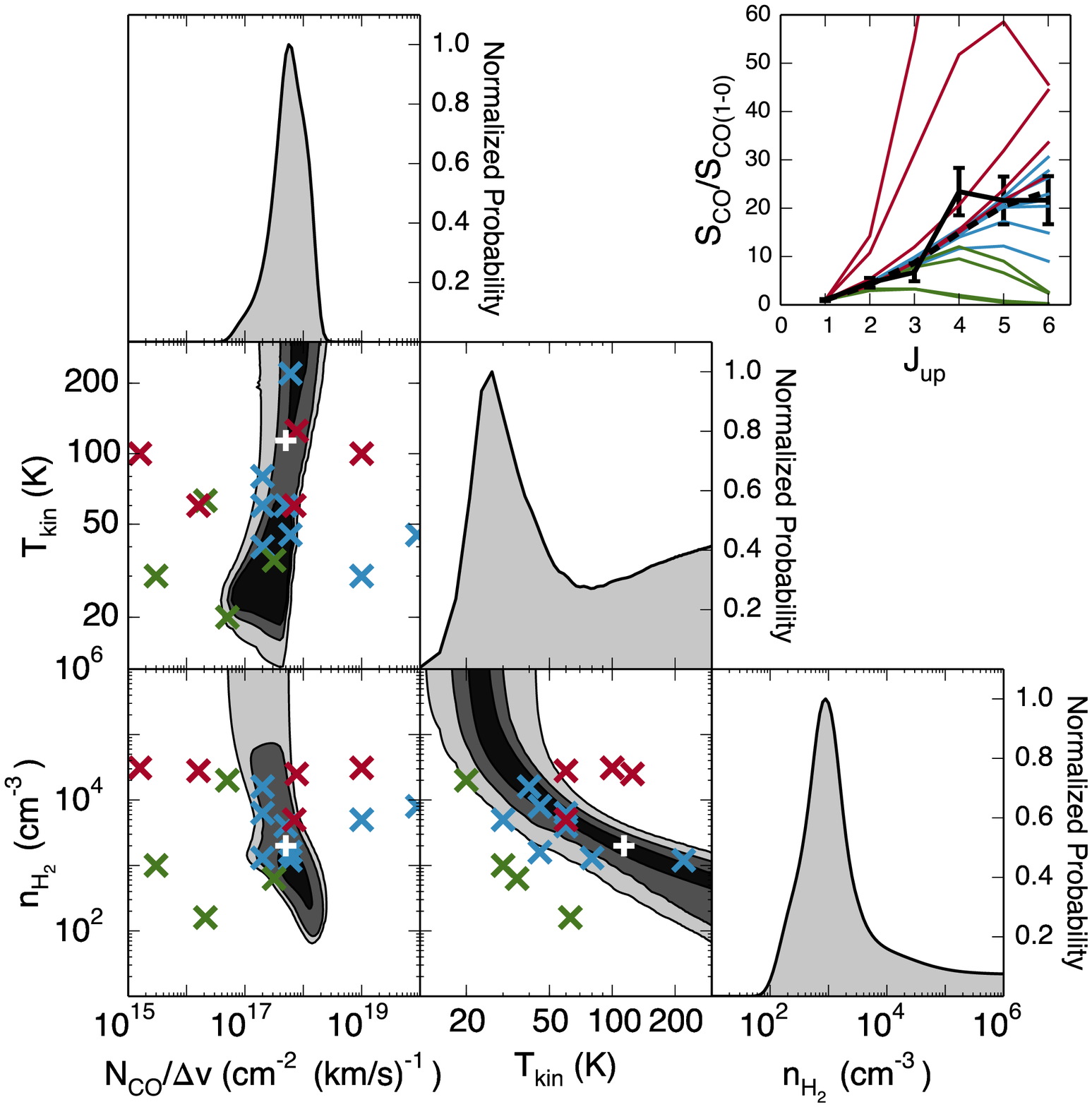}
\caption{
\textit{Left: }Parameter degeneracies in the recovery of a single set of input conditions.  The input conditions are marked in each two-dimensional marginalization as an `X.'  As before, contours are 1, 2 and 3$\sigma$. The upper right panel shows the modeled SLED for the input conditions (solid black line with error bars), and the best-fit model SLED (black dashed line). Unsurprisingly, the best-fit model SLED is essentially identical to the model SLED.  The conditions corresponding to this best-fit SLED are shown with a white `+' symbol in the marginalized parameter distributions.  \textit{Right: }Parameter degeneracies when fitting the average spectrum derived from objects with a wide range of conditions, using a stacking procedure similar to that used for the SPT DSFGs themselves.  Objects with QSO-like SLEDs are colored red, DSFG-like SLEDs are blue, and Milky Way-like SLEDs are shown in green.  The upper right panel includes the SLEDs of all the objects used to create the average SLED (colored lines), in addition to the average spectrum (solid black line with error bars) and best-fit model SLED (black dashed line).  Again, the conditions corresponding to the best-fit SLED are shown with a white `+' symbol.
}
\label{fig:stacktest}
\end{figure}

We begin by attempting to recover the conditions of a single gas component by generating line ratios for a given set of parameters, with no additional scaling, weighting, or averaging.  This test shows the degeneracy that is inherent to the modeling itself -- certain combinations of temperature and column and number densities produce line ratios which are virtually indistinguishable.  In particular, we recover the well-known degeneracy between gas kinetic temperature and number density of molecular hydrogen (Figure \ref{fig:stacktest}, left).  Without additional information -- either in the form of priors on the parameters or additional \twco lines -- this degeneracy is virtually impossible to break.  

We next demonstrate a more realistic scenario, in which the averaging performed in this work is applied to galaxies with a wide range of gas conditions.  In this test, we generate SLEDs based on gas conditions reported in the literature that were determined with similar modeling of \twco lines.  We include a variety of objects from the literature, which can generally be placed in one of three categories: Milky Way-like -- relatively quiescent galaxies with \twco SLEDs that peak at $J < 5$; DSFGs and ULIRGs -- galaxies with active star formation and SLEDs peaking at $5 \leq J < 7$; and QSOs -- galaxies in which heating in the nuclear regions drives the SLEDs to peak at $J > 7$.  For each of 22 SPT DSFGs, we randomly assign the galaxy to be represented by a single gas component model drawn from literature objects.  We add normally-distributed noise to each line in accordance with the typical S/N of our observed \twco lines (median S/N $\sim$8).  Finally, we calculate the average SLED, using only those lines from each object that were observed in our sample (such that, for example, the CO(1-0) line is an average of five input objects).  The results of this test are shown in Figure \ref{fig:stacktest}, right.  While avoiding the term ``average'' due to its natural mathematical connotation, conditions representative of the inputs are recovered despite the variety of conditions used to generate the input spectra.  Conditions are recovered similarly well when the averaging is performed on each of the three categories of sources individually.  The results of this test indicate that the uncertainty in determining gas conditions is largely dominated by the systematic uncertainty and degeneracies introduced by the radiative transfer modeling itself, rather than from the scatter of the conditions which produce the variety of observed \twco SLEDs.

\end{document}